\documentclass[12pt,a4paper]{article}

\usepackage{lineno}
\usepackage{svg}
\usepackage{siunitx}
\usepackage{authblk} 
\usepackage[numbers]{natbib}

\usepackage{amsmath, amssymb, amsfonts}
\usepackage{bm}
\usepackage{mathtools}
\usepackage{upgreek}   
\usepackage{siunitx}   

\title{On the role of water activity on the formation of a protein-rich coffee ring in an evaporating multicomponent drop}

\author[1]{Javier Martínez-Puig\thanks{Corresponding author: jmpuig@pa.uc3m.es}}
\author[2]{Gianluca D'Agostino}
\author[2]{Ana Oña}
\author[1]{Javier Rodríguez-Rodríguez}

\affil[1]{Carlos III University of Madrid, Department of Thermal and Fluids Engineering, Leganés, 28911, Spain}
\affil[2]{Centro Nacional de Biotecnología, Consejo Superior de Investigaciones Científicas CSIC, Madrid, Spain}

\date{} 

\begin{document}
\maketitle

\begin{abstract}

The coffee-ring effect is a universal feature of evaporating sessile droplets with pinned contact line, wherein solutes or particles are advected to the droplet’s edge due to evaporation-driven flows. While existing models have successfully described this phenomenon in particle-laden droplets, they often assume that the evaporative flux, and thus hydrodynamics, are decoupled from solute transport. This assumption breaks down in complex fluids, such as protein or polymeric solutions, where the solute can influence evaporation through changes in water activity. Here, we investigate model respiratory droplets primarily composed of water, salt, and a type of the glycoprotein mucin. Using fluorescence microscopy, we observe the formation of a well-defined protein ring at the droplet edge as water evaporates. The growth and morphology of this ring exhibit a strong dependence on ambient relative humidity ($H_r$), revealing dynamics that existing models cannot capture. Specifically, we find that protein accumulation at the edge is governed by the feedback between local solute concentration and evaporation rate. To account for this, we develop a minimal theoretical model based on the lubrication approximation, incorporating the coupling between hydrodynamics and solute transport through the evaporation rate. Our framework reproduces key features of the experimental observations and suggests a physical basis for the $H_r$-dependent stability and infectivity of respiratory droplets containing viruses.

\end{abstract}


\section{Introduction}\label{sec:Intro}

Since the seminal works by the Chicago group \citep{deegan1997cofrin, deegan2000contact}, the coffee-ring effect (CRE) has garnered significant attention (\cite{WilsonAmbrosio}). In an evaporating sessile droplet, the pinning of the contact line imposes a radial outward flow to enforce mass conservation, transporting material from the center toward the edge. In particle-laden droplets, this advective flow carries particles to the contact line, resulting in the characteristic ring-shaped deposit that has been extensively studied through simplified models of the CRE. However, in more complex fluids, this transport involves not particles but solutes such as polymers, proteins, or salts. The evaporation of such complex fluids is relevant in many situations of practical interest (see for example, \cite{DoiFluoTech, GuoDexCRE, MailleurCRESalt, BasuCRESalt}). Despite its relevance, the physics behind CRE formation in complex droplets remains less thoroughly explored.

In this work, even though similar problems arise in other complex fluids, we focus on protein ring formation in sessile model respiratory droplets. Since the COVID-19 pandemic, this topic has gained significant attention in the disease transmission science community due to its implications for viral infectivity. Remarkably, viral particles (virions) can remain infectious for hours within the dry residue left by evaporated respiratory droplets \citep{morris2021mechanistic, merhi2022assessing, oswin2022dynamics}. This is unexpected, as complete evaporation leads to high salt concentrations which creates, presumably, a harsh chemical environment for virions \citep{seyfert2022stability}. One possible explanation to this paradox is that the droplet components—primarily salt and protein \citep{vejerano2018physico}—segregate during evaporation. A leading hypothesis, yet to be confirmed \citep{morris2021mechanistic, martinezpuig2025}, is that virions are trapped in protein-rich regions, which protect them from salt-induced damage. Understanding how the protein ring forms is therefore needed to clarify how viral infectivity depends on relative humidity, $H_r$.

Some recent studies have addressed this problem qualitatively from a biological perspective. For example, \cite{kong2022virus} and \cite{pan2025mucin} analyzed the final residue of evaporated sessile respiratory droplets and found that the final width of the protein ring depends on relative humidity —a behavior that contrasts with predictions for particle-laden droplets. Another intriguing observation is a transition above $H_r = 80\%$: at lower humidity, a protein-rich ring forms near the contact line, while at higher humidity protein is more uniformly distributed, without forming any distinctive structure. Despite these qualitative differences from the classical coffee-ring effect in particle-laden systems, a dedicated theoretical framework for the ring formation in evaporating complex droplets—such as those containing both salt and protein—has not yet been developed.

A landmark mathematical analysis of the dynamics of deposition rings in particle-laden droplets was carried out by \cite{Popov}, whose framework has since been widely adopted. This model, developed specifically for particle suspensions, relies on two key assumptions. First, the evaporation rate is assumed to be decoupled from the solute transport problem, so it can be readily computed by solving the diffusion-driven vapor transport outside the drop for a constant vapor concentration at the air-droplet interface. Then, the evaporation rate thus obtained is used to compute the advective flux in the transport equation. Second, the formation of the deposition ring is driven by this advective flux, which transports particles toward the contact line. As the local particle concentration approaches the maximum packing fraction, further incoming particles cannot penetrate the already-deposited ring and instead accumulate at its inner edge, causing the ring to broaden towards the drop center.

More sophisticated models, such as the one developed by \cite{sprittles} to study the so-called surface capture, build upon these principles. These hypotheses, reasonable for particle-laden suspensions, are crucial for solving the problem analytically as they enable the decoupling of hydrodynamic and transport equations. A recent, very elegant, study by \cite{moore} introduced the effect of particle diffusion to investigate the early stages of the ring formation using asymptotic techniques. They solve an advection–diffusion equation for the particle concentration without imposing a maximum particle packing fraction. As a consequence, the particle concentration can grow unbounded, which leads to a progressive narrowing of the deposit, with the ring width decreasing in time and collapsing to a single line as 
$t\to t_{ev}$. This behaviour is not consistent with experimental observations, which show a finite ring thickness at all times, including at the drying time. For this reason, most models aiming at quantitative comparison with experiments incorporate a maximum particle packing fraction.

We refer to these models as constant-activity models (CAM). Although they align well with experiments on particle-laden droplets—for instance, \citet{coombs2024colloidal} successfully reproduced the measurements of \citet{LiRateDep}—the concept of a particle packing fraction does not naturally extend to complex droplets. Moreover, these theories fail to capture several behaviors observed in polymeric or protein solutions. For example, studies have shown that in evaporating model respiratory droplets, the width of protein rings formed via the coffee-ring effect depends on relative humidity ($H_r$) \citep{kong2022virus, pan2025mucin}. While this dependence may appear intuitive to experimentalists, it contradicts existing reduced models of the coffee-ring effect. In Popov’s framework, $H_r$ only enters through the overall evaporation timescale, implying that the final deposition ring width should be independent of $H_r$.

In contrast to particle-laden systems, where ring formation is commonly explained by the existence a maximum packing fraction, in polymeric or protein-rich droplets a more natural explanation involves water activity, $\chi_w$. This thermodynamic property quantifies the water vapor pressure at an interface of an aqueous solution relative to that pure water, i.e $C_{\mathrm{vap}} = C_s\, \chi_w$, where $C_{\mathrm{vap}}$ is the concentration of water vapor at the air-droplet interface and $C_s$ is the water vapor saturation concentration in air. 
The water activity takes into account how solutes (like proteins or salts) reduce the evaporation rate. 
For solutes with a large affinity for water, such as some salts, the water activity can be much smaller than unity, which impacts strongly the evaporation dynamics. For an ideal mixture with solute mass fraction $w_p$, the water activity is given by $\chi_w = 1 - w_p$. For more complex solutions that depart from ideality, we still have, in general, $\chi_w \to 0$ as $w_p \to 1$; that is, there is no water vapor at the interface of the droplet if there is no water inside the droplet at that location.

To our knowledge, \cite{RogerK_ContrWatEvp} were the first to experimentally study the qualitative changes in evaporation behavior in terms of water activity. Among their findings for an evaporating capillary is that, for a simple binary mixture, the evaporation rate becomes independent of ambient relative humidity for $H_r \leq 85\%$. \cite{Salmon} studied theoretically this surprising phenomenon. In their work on one-dimensional evaporation of supramolecular mixtures, they showed that evaporation can become nearly insensitive to ambient humidity. This occurs when the water activity drops sharply at high solute concentrations, so that the solute concentration at the drying interface becomes nearly independent of $H_r$, leading to evaporation rates that vary only weakly with humidity. \cite{HuismanPRL} experimentally verified this theory by studying the evaporation of polyvinyl alcohol (PVA)–water solutions from open-ended capillaries. They observed that accumulation of PVA at the evaporating interface forms a polarization layer, which affects both the local diffusivity and the water activity. As a result, the evaporation rate becomes nearly independent of ambient relative humidity for $H_r \leq 80\%$.

Recently, \cite{Raju} investigated the evaporation of glycerol–water solutions from open-ended capillary tubes using both experiments and theoretical modeling. By tracking the height of the upper meniscus, which is not exposed to evaporation, they identified an initial linear decrease in height over time—corresponding to the so-called constant-evaporation-rate stage. As evaporation progressed, glycerol concentration increased at the interface, modifying the water activity $\chi_{w}$ and triggering a transition to the falling-rate stage, where the meniscus velocity followed a power-law decay in time with an exponent of $-1/2$, and a pre-factor that depends on the relative humidity and the initial glycerol concentration. For this same system, \cite{MilarkSalmonSobacPRF2026} have explored the evaporation of a cylindrical drop confined in a Hele-Shaw cell. In this study, the spatio-temporal evolution of the water and glycerol concentrations is accurately measured using Mach-Zehnder interferometry. This allows the authors to determine the mutual diffusion coefficient between water and glycerol as a function of the mixture composition.  Using the experimentally determined diffusion coefficient, the authors reproduce all the results modeling both the solute transport inside the drop and the vapor transport outside as purely diffusive, proving the ability of this kind of models to describe evaporation in confined systems once the water activity and the correct value of the diffusion coefficient are incorporated into the formulation.

A number of studies have also focused on the role of water activity in the evaporation and stability of respiratory droplets. For example, \cite{merhi2022assessing} combined experimental measurements with theoretical modeling to highlight that incorporating a non-ideal thermodynamic water activity establishes an equilibrium droplet size that becomes independent of relative humidity. \cite{seyfert2022stability} investigated the evaporation of model respiratory droplets on superhydrophobic substrates. They found that at high relative humidity, evaporation halts once the water activity of the solution equals the ambient humidity, i.e., $\chi_{w}(w_s) = H_r$. Notably, this condition is reached before salt efflorescence occurs (at 20°C for $H_r \geq 0.75$), resulting in droplets that remain stable and liquid indefinitely without fully drying. More recently, \cite{HautRespDropDyn} studied theoretically how the effects of a non-constant water activity impact both the sedimentation and drying times of spherical respiratory droplets with radius smaller than 50~$\mu\mathrm{m}$. Their findings are in agreement with the experimental results of \cite{merhi2022assessing}, which show that the equilibrium droplet radius is independent of relative humidity for values below approximately 70\%.

These examples illustrate a broader class of systems where evaporation rates are strongly influenced by the composition at the evaporating interface. This effect must be incorporated into any theoretical framework aiming to describe the coffee-ring effect in complex droplets. As we will demonstrate for the case of model respiratory droplets, including the dependence of the water activity coefficient, $\chi_w$, on the protein mass fraction, $w_p$, allows us to explain key experimental observations, such as the relative humidity dependence of the time evolution, and final value, of the ring width.

Our objective is to integrate recent advances from these studies into a comprehensive understanding of coffee-ring formation in complex droplets, with a particular focus on model respiratory droplets as a case study. To this end, Section \ref{sec:Experiments} presents experimental observations of protein ring formation using fluorescence microscopy. In Section \ref{sec:RedModel}, we derive a minimal model that captures the key mechanisms driving the coffee-ring effect in these complex solutions. Section \ref{sec:Results} details the formation mechanism of the protein ring, discusses the roles of salt and protein-dependent diffusivity, compares the theoretical prediction with our experimental results and presents some limitations of the model. Finally, the manuscript concludes with a summary and outlook in Section \ref{sec:Discussion}.

\section{Experiments}\label{sec:Experiments}

We evaporate artificial saliva droplets following the composition used by \cite{vejerano2018physico}, namely milliQ water with \SI{9}{\gram\per\liter} of NaCl (Sigma-Aldrich), \SI{3}{\gram\per\liter} of porcine gastric mucin type III (Sigma-Aldrich) and \SI{0.5}{\gram\per\liter} of the pulmonar surfactant 1,2-dihex-adecanoyl-sn-glycero-3-phosphocholine. Mucin, the main protein found in respiratory fluids, will be particularly important here, as it is the solute responsible for the ring formation. This composition is commonly used in experimental studies investigating the decay rates of viral infectivity \citep{pan2025mucin}.

\begin{figure}[]
    \centering
    \includegraphics[width=0.7\textwidth]{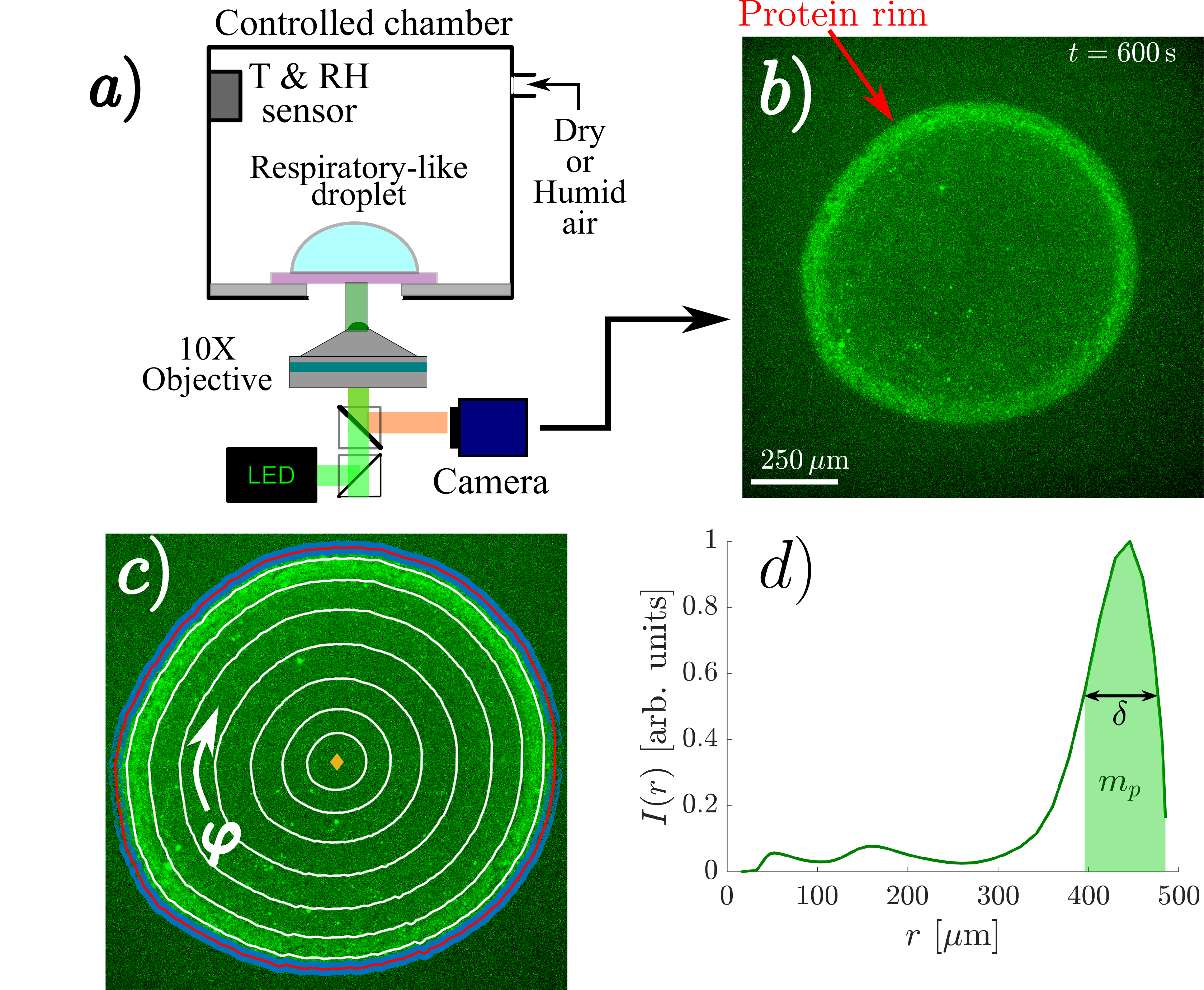}
    \caption{Schematic of the procedure used to obtain the radial intensity profile. a) Time-lapse images of the evaporating droplet are acquired using an epifluorescence microscope inside a humidity-controlled chamber. b) Example of a fluorescence image obtained during evaporation. c) The droplet is segmented into concentric annular regions, and the mean fluorescence intensity is computed for each region. For visualization purposes, fewer regions are shown than were used in the actual analysis. d) From this process, the radial intensity profile is extracted for each image.}
    \label{fig:ExpSetup}
\end{figure}

\subsection{Experimental setup}

We deposit sessile model respiratory droplets of initial volume $V_0=0.3\pm 0.1$ \SI{}{\micro\liter} on $\mu$-Dish 35 mm, high glass-bottom dishes (Ibidi). These micropetri dishes enable high-resolution imaging and do not exhibit autofluorescence. The evaporation follows a constant-radius mode, which is consistent across all experiments and throughout the droplet lifetime. This behavior results from the accumulation of mucin at the substrate, which pins the contact line. The droplets are evaporated inside a humidity-controlled chamber, at a constant relative humidity ($H_r$) which varied between $30\%$ and $70\%$ among different experiments. We use the temperature and humidity sensor DHT11 with an accuracy $\pm 5 \%$ in relative humidity and $\pm 1^\circ$C in temperature. The room temperature varies among experiments between $20^\circ$C$-28^\circ$C. To increase humidity, we introduce humid air into the chamber by recirculating it through a water bottle. For desiccating air, we recirculate it through a separate bottle containing calcium phosphate, a highly hygroscopic salt. This configuration is adapted from \cite{BoulogneHumSetup} in order to fit in the reduced space available on a microscope. An scheme of our experimental experimental setup is shown in figure \ref{fig:ExpSetup}. 

To evaluate the spatial distribution of proteins, we utilized the enhanced autofluorescence of mucin, a highly glycosylated protein abundant in model respiratory droplets. Mucin exhibits high autofluorescence compared to other proteins due to its extensive glycosylation, which can generate fluorescent compounds through sugar oxidation or glycation. Evaporation dynamics were monitored using an inverted widefield fluorescence microscope (Leica DMi8) equipped with a CoolLED pE-4000 illumination system and an Orca-Flash 4.0 sCMOS camera (Hamamatsu). Images were acquired using a 10×/0.32 HC PL FLUO objective. Excitation was performed using a 405 nm LED. The microscope was equipped with a quad-band fluorescence filter set (QUA-T), however, only the channel corresponding to excitation at 405 nm and emission collection at 420–480 nm was used in this study. 
This wavelength is the closest to the UV range, where optimal excitation for proteins occurs, that we have available in our microscope. Despite not being able to excite the protein fluorescence emission at its optimal wavelength, still enough autofluorescent signal is emitted when illuminating in the UV-blue range and imaging the emission in the blue band \cite{lakowicz2006principles}.
The exposure time was set to 20 ms per acquisition. To prevent excessive heating of the droplet, we capture images every 5 seconds and verify that the evaporation time matches the theoretical predictions described by \cite{seyfert2022stability}. 

Due to the axisymmetric nature of the protein rim formation, we focus on the radial intensity profile, defined as:
\begin{equation}
    I(r) = \frac{1}{2\pi} \int_0^{2\pi} i(r,\varphi)\, d\varphi,
\end{equation}
where $i(r,\varphi)$ is the local intensity in polar coordinates. As the droplet's contact line is not a perfect circle, the radial intensity $I(r)$ is estimated through the following procedure. First, the contact line is extracted from the image and represented as a spline. The centroid is then calculated using $A = \frac{1}{2} \sum_{i=1}^{n} (x_i y_{i+1} - x_{i+1} y_i)$, $x_c = \frac{1}{6A} \sum_{i=1}^{n} (x_i + x_{i+1})(x_i y_{i+1} - x_{i+1} y_i)$, and $y_c = \frac{1}{6A} \sum_{i=1}^{n} (y_i + y_{i+1})(x_i y_{i+1} - x_{i+1} y_i)$, where $(x_c, y_c)$ is the centroid of the spline defined by the ordered vertices $(x_i, y_i)$.
 The droplet is then divided into annular regions by constructing inward homothetic splines of the contact line (figure \ref{fig:ExpSetup}c). Within each of these regions, the local intensity is integrated to obtain the corresponding radial intensity. Each homothetic spline is assigned a radius, defined as the mean distance of its points to the centroid of the original contact line. The radial position associated with each annular region is taken as the average of the radii of its inner and outer bounding splines.

\begin{figure}[]
    \centering
    \includegraphics[width=\textwidth]{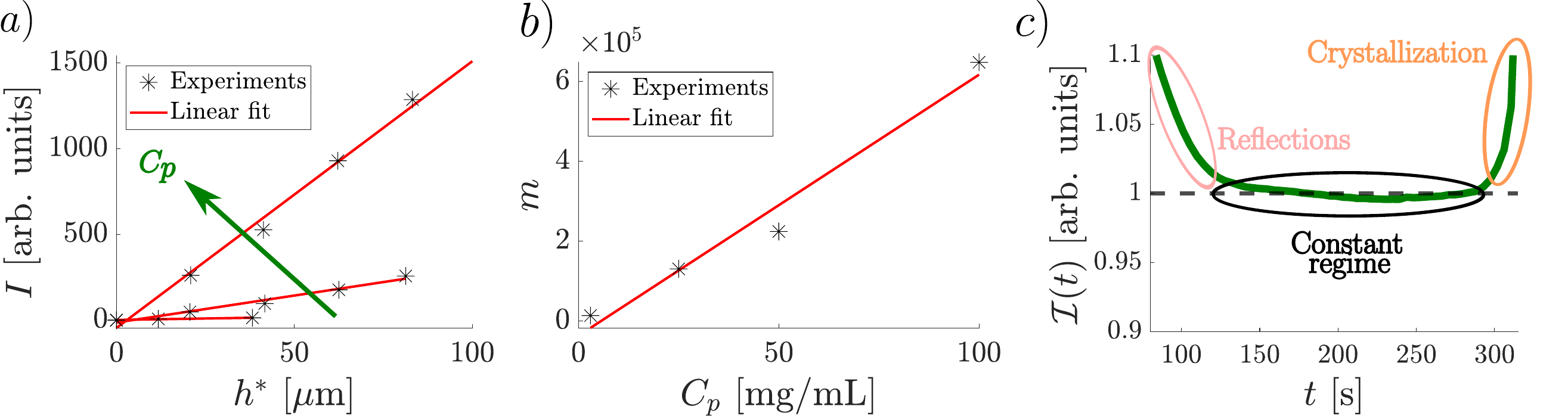}
    \caption{Calibration of fluorescence intensity as a function of droplet height and protein mass fraction. (a) Fluorescence intensity profiles at different relative heights $h^*$ for varying protein mass fraction $C_p = 0.003, 0.025, 0.1$, confirming the linear relationship between intensity and height. (b) Slopes of the fitted lines, $m$, from (a) demonstrating the linear dependence of fluorescence intensity on protein mass fraction. (c) Representative time evolution of the total fluorescence intensity, showing distinct regimes during evaporation. All quantitative analyses presented throughout this article are based on data from the constant-intensity regime.}
    \label{fig:Cal_ConstInt}
\end{figure}

In order to obtain the height-averaged mass fraction of mucin, $\overline{w}_p$ from the radial intensity $I(r)$ we follow a similar approach to that employed by \cite{DoiFluoTech}. We assume that the fluorescence intensity follows the relationship  
\begin{equation}\label{eq:LinearIr}
I(r) = K \overline{w}_p(r) h(r),
\end{equation}
where $K$ is a proportionality constant, and $h(r)$ is the height of the drop. To verify the linear relationship between fluorescence intensity, height-integrated protein mass fraction, and droplet height, we performed a series of calibration experiments. A calibration chamber was constructed using glass slides arranged in an L-shaped geometry, where $L$ denotes the horizontal length of the base and $H$ the vertical height. A thin coverslip was placed on top to enclose the chamber. This configuration creates a wedge-shaped volume of solution with a linearly varying height $h = x (H/L)$ where $x$ is the horizontal direction. By translating the microscope objective along this direction, we can access fluorescence signals at known and controllable sample heights, enabling calibration of the fluorescence response as a function of height. We acquired fluorescence images of protein solutions with $\overline{w}_p = 0.003,\, 0.025,\, 0.05,\, 0.1$. Higher protein mass fraction ($\overline{w}_p > 0.1$) could not be dissolved in water and salt to produce a uniform mixture. For each protein mass fraction, we verified the linear relationship between fluorescence intensity and solution height (see figure~\ref{fig:Cal_ConstInt}a). To test the dependence on $\overline{w}_p$, we found that the slopes of the fitted lines relating intensity to height scale linearly with the height-integrated protein mass fraction (see figure~\ref{fig:Cal_ConstInt}b). As a consequence of this linearity, from equation \eqref{eq:LinearIr}, the total amount of light intensity $\mathcal{I}(t)$ follows

\begin{equation}
    \mathcal{I}(t)=\int_0^{R_c}2\pi r I(r,t)dr=\int_0^{R_c}2\pi r K \overline{w}_p(r,t) h(r,t)dr= K M_p,
\end{equation}
where $R_c$ is the contact radius and $M_p$ is the total amount of protein in the droplet, that remains constant throughout the experiment. At the beginning of the evaporation process, the relatively high contact angle (approximately $60^\circ$) leads to optical reflections near the contact line, which artificially increase the measured fluorescence intensity $\mathcal{I}(t)$. To avoid this artifact, our analysis begins only after these reflections disappear and the system enters a regime in which $\mathcal{I}(t)$ remains constant. Within this regime, the intensity varies by less than 1\% throughout each experiment. Following this stable period, droplets exposed to low and medium relative humidity eventually crystallize, causing a sharp increase in $\mathcal{I}(t)$. Our analysis is therefore restricted to the temporal window between the disappearance of initial reflections and the onset of crystallization (see figure~\ref{fig:Cal_ConstInt} c). 

It should be noted that although both mucin and salt crystals were visualized using fluorescence microscopy with excitation at 405 nm and emission detection between 420–480 nm, only mucin exhibits intrinsic autofluorescence in this spectral window. Inorganic salt crystals do not emit fluorescence under these conditions. The signal observed from salt crystals is therefore attributed primarily to light scattering rather than true fluorescence emission. Additionally, it is possible that mucin molecules become adsorbed or deposited onto the surfaces of salt crystals during droplet evaporation, resulting in localized fluorescence signals that spatially coincide with the crystal structures. A detailed investigation of this complex interplay is beyond the scope of the present work, but it constitutes an interesting avenue for future research.

From fluorescence experiments, we obtain the radial intensity as $ I(r) = K \, \overline{w}_p \, h $. However, due to spatial constraints within the microscope setup, it is not possible to image the droplet from the side during fluorescence experiments. As a result, we cannot directly measure the interface height profile $ h(r,t)$. To circumvent this limitation, we work with the height-integrated protein fraction $\phi_p = h\,\overline{w}_p$ for analyzing the spatial distribution of protein.

\subsection{Experimental results}

\begin{figure}[ht!]
    \centering
    \includegraphics[width=\textwidth]{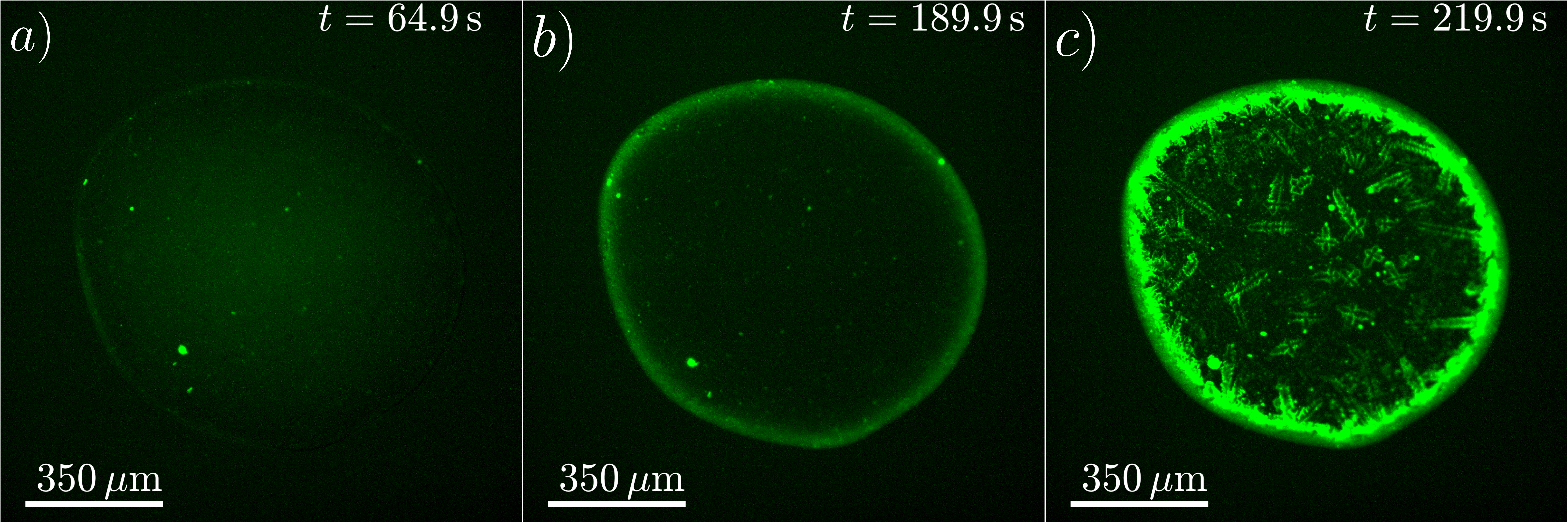}
    \caption{Fluorescence images showing a) the initial homogeneous distribution of protein, b) the formation of a peripheral protein ring, and c) final crystallization. Droplet evaporated at $H_r=30.4\%$ and $T=20.5^\circ$C.}
    \label{fig:ResExpImages}
\end{figure}

As evaporation progresses, protein begins to accumulate near the contact line, forming a protein ring. We observe that this protein ring increases in both intensity and width over time (see figure~\ref{fig:ResExpImages}). Notably, in model respiratory droplets, a relevant feature of the protein ring is its width, $\delta$, which has been linked to the low decay rates of virions’ infectivity in dry residues of model respiratory droplets \citep{HuangWamgVikeslandLANGMUIR2021}. Protein is expected to shroud virions, slowing their loss of infectivity \citep{wardzala}. This suggests that a wider protein-rich region could help virions to remain infective for a longer period.  

We define the width of the coffee-ring, $\delta$, as that of the peak at half prominence (figure~\ref{fig:ExpSetup} d). We adopt this definition because measuring the peak width at half height, as done in analytical theories like those of \cite{Popov} and \cite{moore}, is experimentally challenging for high relative humidity. In these cases, the height of the peak does not allow for easy identification of the intersection at half height in the inner part of the protein ring.

\begin{figure}[ht!]
    \centering
    \includegraphics[width=\textwidth]{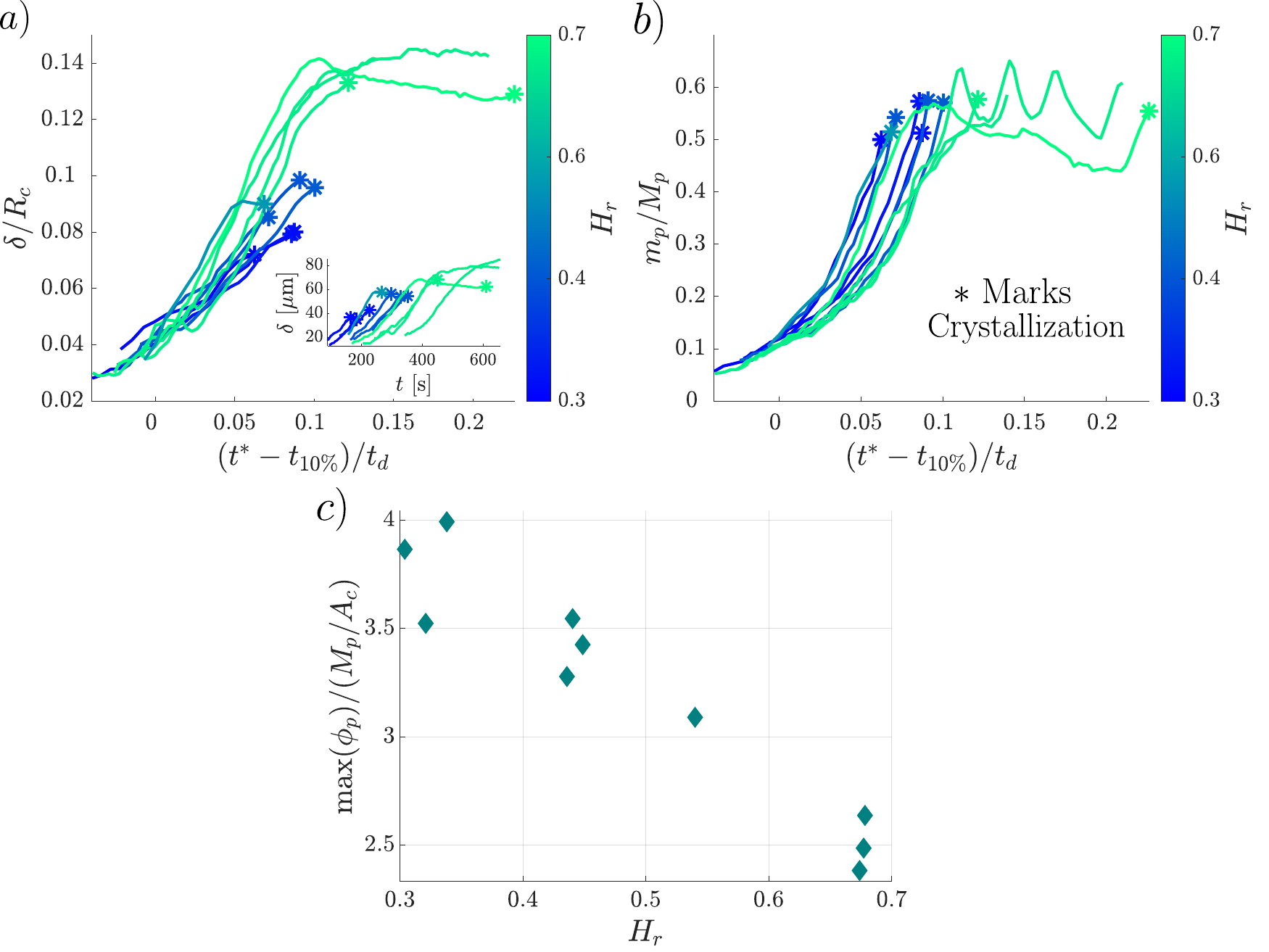}
    \caption{Experimental results. (a) Temporal evolution of the ring width, $\delta$. (b) Mass of protein, $m_p$, in the ring normalized by the total protein mass in the droplet,  $M_p$, as a function of time. (c) Dependence of the maximum height-integrated protein fraction, normalized by the total protein mass in the droplet divided by the contact area $A_c$, with the relative humidity.}
    \label{fig:ResExp}
\end{figure}

The most striking phenomena we found in experiments is that the width $\delta$ follow different dynamics for different relative humidity (see figure~\ref{fig:ResExp} a and appendix~\ref{app:Exp_Residue_by_Hr}) and that the maximum height-integrated protein fraction $\phi_p$ of the protein ring decreases as the relative humidity increases (see figure~\ref{fig:ResExp} c and appendix~\ref{app:Exp_Residue_by_Hr}). As explained on the introduction in constant-activity models the width of the coffee ring has been found to follow the same scaling law across all values of the relative humidity. However, as shown in figure~\ref{fig:ResExp}(a), this behavior does not hold for protein ring formation in sessile model respiratory droplets. In this figure we show the time evolution of the rim width for drops evaporating at different humidity, with the time scaled with the diffusive timescale, $t_d = \rho R_c^2 / D_{\text{vap}} C_{\text{s}}(1-H_r)$. This timescale is computed with the density of water, $\rho$, the contact radius, $R_c$, the diffusivity of water vapor on air, $D_{\text{vap}}$, the saturation concentration of water vapor in air, $ C_{\text{s}}$ and the ambient relative humidity, $H_r$. Because of experimental limitations in determining the exact onset of deposition, the initial time of each experiment carries some uncertainty. To account for this, we reset the origin of the temporal axis by introducing $t_{10\%}$, defined as the time at which the protein mass accumulated in the ring reaches $10\%$ of the total protein mass in the droplet.

Figure~\ref{fig:ResExp} a shows that the evolution of the rim width $\delta$ depends on the relative humidity $H_r$ in two distinct ways. First, the final rim width attained prior to crystallization varies with $H_r$, with larger values observed at higher relative humidity; this dependence is absent from the model of \citet{Popov}, in which all curves collapse when time is rescaled by the diffusive timescale $t_d$. Second, the growth rate of the rim width differs between humidities: when time is non-dimensionalized by $t_d$, $\delta$ grows more slowly at low and intermediate $H_r$ than at high $H_r$. The slopes are different over the evolution, and at high relative humidity  $\delta$ saturates at late times. In more sophisticated theories, such as those of \citet{sprittles} and \citet{moore}, the temporal evolution of the ring width exhibits a Péclet dependent offset while retaining a universal slope. In contrast, our experiments reveal an increase of the slope with increasing $H_r$, a feature not captured by existing coffee-ring models. This behavior can be rationalized by accounting for the dependence of water activity on both salt and protein mass fractions (see Section \ref{sec:Results} for a detailed discussion).

\section{Mathematical model}\label{sec:RedModel}

In this section, we present a minimal theoretical model aimed at capturing the essential dynamics of the coffee-ring effect (CRE) in complex droplets, as observed in our experiments (see Section \ref{sec:Experiments}). The goal is not to achieve a one-to-one quantitative comparison with the experimental results, but rather to use the experimental observations as motivation to explore physical mechanisms that are not adequately explained by existing theoretical models.

As discussed in the Introduction, most existing models of the coffee-ring effect (CRE) have been developed in the context of particle-laden droplets. To reproduce the formation of a realistic ring width, these models typically rely on the assumption of a maximum packing fraction $w_{p_c}$ \citep{Popov,Kaplan_Mahadevan_2015, sprittles}. This assumption leads to a separation of the droplet into two distinct regions: one where the local particle concentration is exactly $w_{p_c}$, forming the dense particle ring, and another where the concentration remains below $w_{p_c}$, corresponding to a dilute region. These frameworks allow for a decoupling of hydrodynamics from the transport problem, simplifying the analysis to make it tractable analytically. While this approach is appropriate for particle suspensions, it becomes less adequate when applied to polymeric or protein solutions as, in such systems, there is no natural extension of a maximum packing fraction. As a result, the use of $w_{p_c}$ in this context is somewhat arbitrary in these systems and does not accurately capture the underlying physics of polymer or protein ring formation. A more sophisticated approach, adopted by \cite{coombs2024colloidal}, is to introduce a concentration-dependent viscosity, which effectively models the ring as a region where particles progressively jam until they reach a critical concentration. An advantage of this approach is that it is equally valid for particle suspensions and for solutions of polymers or other complex molecules. In such case, the critical concentration can be, for instance, that at which a gel transition occurs. However, as we will discuss below, this approach is still not able to explain the effect that the ambient relative humidity is observed to have on the drop dry deposit, as evaporation is still decoupled from the transport dynamics.

In contrast, here we adopt a different approach in which the entire droplet is treated as a single continuous phase. Consequently, the equations we derive are valid throughout the entire droplet and remain applicable at all times during the evaporation process.

As it has been shown \citep{martinezpuig2025}, the flow pattern in respiratory droplets can be recirculatory due to natural convection. In other contexts, recirculation may also arise from Marangoni stresses \citep{marin2019solutalmarangoni, dekker2025internalflowconcentrationneighbouring}. For a detailed two-dimensional model that accounts for these phenomena, we refer the interested reader to \citep{Diddens2021JFM, dekker2025internalflowconcentrationneighbouring,martinezpuig2025}. In the present work, however, our goal is to develop a minimal model that explain the mechanism behind the coffee ring effect in complex droplets. To that end, we adopt the lubrication approximation and derive a one-dimensional model to describe the hydrodynamics and solute transport within the droplet.

\subsection{Flow problem}

In order to apply the lubrication approximation, we assume that $\epsilon = h_0 / R_c \ll 1$, where $h_0$ is the initial height of the droplet at $r = 0$. Also, we neglect the effect of gravity. This has two implications: first, we assume a small Bond number $\mathrm{Bo} = \rho g R_c^2 / \gamma^* \ll 1$, where $\gamma^*$ is the surface tension. This hypothesis implies that the shape of the drop's surface is not affected by gravity, and is valid for aqueous droplets with radii smaller than a millimeter. Second, we also neglect buoyancy forces in the hydrodynamic and transport equations. In point of fact, the stratification caused by evaporation can drive a recirculatory buoyancy-driven flow in our system at short times, when the drop is not shallow. This is due to the fact that the ratio between the buoyancy-driven and the capillary flow scales as $\varepsilon^5\mathrm{Ra}$, where $\mathrm{Ra} = \partial_w\rho g R_c^3 / \mu^* D^*_p$ is a Rayleigh number \citep{Diddens2021JFM}. In this definition, $\partial_w\rho$ is the derivative of the solution's density with the solute mass fraction, $\mu^*$ the liquid viscosity, and $D^*_p$ the diffusivity of the solute. For the droplet sizes and compositions of our experiments, $\mathrm{Ra} \sim 10^6$, which generates a recirculatory flow driven by density stratification at short times when $\varepsilon$ is still not small enough \citep{martinezpuig2025} (see also Appendix \ref{app:BuoyancyDrivenFlow}). Nonetheless, we neglect these effects in what follows for two reasons: first, we work with height-averaged radial velocity $\bar{u}_r$ and protein mass fraction $\bar{w}_p$, thus the velocity and concentration profiles do not appear explicitly in our formulation. Second, the buoyancy-driven recirculatory flow does not transport any net mass along the radial direction, for the reasons pointed out by \cite{Diddens2021JFM}, so its effect on the solute transport can be neglected in a first approximation.

Marangoni effects can also be incorporated within the lubrication framework. However, previous studies \citep{Du2022, martinezpuig2025}, have shown that surface tension gradients are negligible in protein-laden mixtures, due to the accumulation and jamming of proteins at the interface. Thus, we neglect surface tension variations on the model. We further assume that the flow is axisymmetric so we have, in cylindrical coordinates:

\begin{equation}\label{eq4:DimensionalModel}
\frac{\partial h^*}{\partial t^*}+\frac{1}{r^*} \frac{\partial}{\partial r^*}\left(r^* h^* \bar{u}_r^*\right)=-\frac{J^*}{\rho^*}, \quad \bar{u}_r^*=-\frac{h^{* 2}}{3 \mu^*} \frac{\partial p^*}{\partial r^*}, \quad p^*=p_{a t m}^*-\frac{\gamma^*}{r^*} \frac{\partial}{\partial r^*}\left(r^* \frac{\partial h^*}{\partial r^*}\right),
\end{equation}
where $h$ is the height of the droplet, $t$ is the time, $r$ is the radial coordinate, $J$ is the local evaporation rate, $\rho$ is the density of the suspension, $\bar{u}_r$ is the height-averaged radial velocity, $\mu$ is the suspension viscosity, $p$ is the pressure, $p_{atm}$ is the atmospheric pressure and $\gamma$ is the surface tension of the suspension. The superscript $^*$ denotes dimensional variables. Equation \eqref{eq4:DimensionalModel} is subject to symmetry boundary conditions at the symmetry axis

\begin{equation*}
    \frac{\partial h^*}{\partial r^*}=Q^* = 0  \  \mathrm{at} \ r^*=0, 
\end{equation*}
where $Q^* = r^* h^* \bar{u}_r^*$ is the local flow rate. At the contact line there is no flux and the height is zero 

\begin{equation*}
    h^*=Q^* = 0  \  \mathrm{at} \ r^*=R_c.
\end{equation*}
As we have neglected the influence of gravity in the droplet shape the initial condition is that of a paraboloid of revolution (equivalently to a spherical cap in the lubrication approximation).

\begin{equation*}
    h^*= \frac{h_0}{R_c^2}(R_c^2-r^*{^2}) \  \mathrm{at} \ t^*=0.
\end{equation*}

\subsection{Solute transport problem}

We derive a transport equation for the height-averaged solute mass fraction $\bar{w}_p$ integrating vertically from $z^* = 0$ to $z^* = h^*(r^*, t)$ the two-dimensional, axisymmetric transport equation. Doing so, we obtain:
\begin{equation}\label{eq4:SoluteDimensional}
\frac{\partial}{\partial t^*}\left(h^* \overline{w}_p\right)+\frac{1}{r^*} \frac{\partial}{\partial r^*}\left(r^* h^* \bar{u}_r^* \overline{w}_p\right)=\frac{1}{r^*} \frac{\partial}{\partial r^*}\left(D_p^* r^* h^* \frac{\partial \overline{w}_p}{\partial r^*}\right).
\end{equation}
This equation is subjected to symmetry conditions at the axis $r=0$

\begin{equation}
    \frac{\partial \overline{w}_p}{\partial r^*} = 0  \  \mathrm{at} \ r^*=0.
\end{equation}
Moreover, as the solute is non-volatile, we impose a no-flux condition at the contact line  

\begin{equation}
    r^* h^* \bar{u}_r^* \overline{w}_p-D_p^* r^* h^* \frac{\partial \overline{w}_p}{\partial r^*}=0  \  \mathrm{at} \ r^*=R_c.
    \label{eq:BoundaryConditionTransportContactLine}
\end{equation}
We further assume that the solute is initially well-mixed so the initial condition is 

\begin{equation}
    \overline{w}_p = w_{p_0} \  \mathrm{at} \ t^*=0,
\end{equation}
where $w_{p_0}$ is the initial protein mass fraction.

To remain consistent with the equations derived in the previous subsection, we also adopt here the lubrication approximation, $\epsilon\ll1$. Also, we regard the solute as well mixed along the $z$ direction. This amounts to substitute $\overline{u_r^* w_p}$ by $\bar{u}_r^*\bar{w}_p$ in Equations (\ref{eq4:SoluteDimensional}) and (\ref{eq:BoundaryConditionTransportContactLine}). The solute can be considered well mixed if a vertical Péclet number is small, so diffusion is able to smear the stratification caused by the relative motion between the liquid and the receding interface. As explained in \cite{ramirez2022taylor}, the vertical Péclet number is of the order $\mathrm{Pe}_z \sim \epsilon^2\mathrm{Pe} = \epsilon^2 \bar{u}^*_r R_c / D^*_p$. In this expression, one factor $\epsilon$ arises from projecting the radial velocity onto the direction perpendicular to the interface and the other from the fact that diffusion occurs mainly along that same direction.

As will be shown in section \ref{sec:Results}, we get values of order $\mathrm{Pe} \sim {\cal O}(10)$ in our experiments. Consequently, the vertical Péclet number is larger than unity until $\epsilon$ does not decrease below $\epsilon \approx 0.3$. This occurs for a contact angle about $\theta \approx 30^\circ$. Thus, at short times, the solute could not be regarded at well mixed. However, we must take into account that the transport dynamics at the rim are local \citep{moore}, since mass accumulates preferentially near the contact line. So, since the drop height is small in the rim region, the local value of $\mathrm{Pe}_z$ is in fact smaller there than the global value. Moreover, a substantial part of the rim mass accumulates at times close to full evaporation, when $\epsilon \ll 1$. All in all, the vertical Péclet number relevant to describe the solute transport towards the rim can be deemed as small, although it is not during the first part of the evaporation process and outside the rim region.


\subsection{Evaporation rate}

Up to this point, if viscosity and diffusivity are assumed to be constant and independent of solute concentration, the hydrodynamic problem is decoupled from the transport of solute.  However, this decoupling breaks down when we account for the fact that evaporation itself depends on local solute concentration. It is well established that the water activity—and therefore the local evaporation rate—is a function of solute concentration for many polymers and proteins. This introduces a critical feedback mechanism: the solute concentration affects the evaporation rate (\cite{Salmon, Raju}), which in turn modifies the flow field and solute transport. In the following we describe how to account for this coupling through the evaporation rate.

The large evaporation timescale, $ t_{\text{ev}} \sim 10^2 \, \mathrm{s} $, compared to the characteristic diffusive timescale of water vapor in the gas phase, $ t_{d_G} = R_c^2 / D_{\text{vap}} \sim 10^{-1} \, \mathrm{s} $, justifies the assumption of quasi-steady diffusion of water vapor in the surrounding air

\begin{equation}\label{eq4:LaplaceDimen}
    {\nabla^2}^\ast C_{vap}^\ast=0.
\end{equation}
Provided that the droplet is thin, the Dirichlet condition at the droplet-air interface is imposed at $z^*=0$

\begin{equation*}
    C_{vap}^\ast = C_s \chi_w \  \mathrm{at} \ z^*=0, \ 0<r^*<R_c.
\end{equation*}
This condition arises from the assumption of thermodynamic equilibrium between the droplet and the surrounding vapor phase. The water activity, $\chi_w$, accounts for the reduction in the equilibrium vapor concentration at the droplet interface due to the presence of solutes, relative to that of a pure water droplet. This condition needs to be complemented with the symmetry condition 

\begin{equation*}
    \frac{\partial C_{vap}^*}{\partial r^*}=0 \ \mathrm{at} \ r^*=0,
\end{equation*}
the no-flux condition on the substrate outside the droplet 

\begin{equation*}
    \frac{\partial C_{vap}^*}{\partial z^*}=0 \ \mathrm{at} \ r^*>R_c,
\end{equation*}
and the far-field condition 

\begin{equation*}
    C_{vap}^*=C_s H_r \ \mathrm{at} \ r^*\to\infty, \ z^*\to\infty.
\end{equation*}
Formally, on a thin droplet where the water vapor transport is diffusive, the evaporation rate is 

\begin{equation*}
    J^*=-D_{vap}^*\frac{\partial C_{vap}^*}{\partial z^*} \  \mathrm{at} \ z^*=0, \ 0<r^*<R_c.
\end{equation*}

This problem can be solved analytically, as shown by \citet{Popov}, for the case of a constant Dirichlet boundary condition at the air–water interface, i.e., when the water activity $\chi_w$ remains constant. However, to capture the rich phenomenology of the evaporation dynamics of polymeric or protein solutions, the water activity must be treated as concentration-dependent, $\chi_w(\overline{w}_p)$, and therefore varies along the droplet radius. This class of problems is classical in potential theory \citep{Copson_1947}. One approach is to apply the Hankel transform, which reduces the problem to a dual integral equation \citep{Noble1958}, although this method is algebraically cumbersome. A more elegant and tractable solution, based on techniques from complex analysis, was presented by \citet{greenElasticity}, yielding the following expression for the evaporation rate:

\begin{equation}
J^*=-D_{vap}C_s\frac{1}{r^*} \frac{\partial}{\partial r^*} \int_{r^*}^{R_c} \frac{r' g(r') \, \mathrm{d} r'}{\sqrt{r^2-r'^2}}
\label{eq:J_vs_integral_g}
\end{equation}
where 
\begin{equation}
g(r')=\frac{2}{\pi} \frac{\mathrm{d}}{\mathrm{d} r'} \int_0^{r'} \frac{r \left(\, \chi_w(r)-H_r\right) \, \mathrm{d}r}{\sqrt{r'^2-r^2}}.
\label{eq:g_vs_integral_chiw}
\end{equation}

Notably, this formulation recovers the analytical solution of \citet{Popov} as a special case when $\chi_w$ is constant. Solving these integrals analytically for a non-constant $\chi_w$ remains far from trivial. \citet{FUDopantConcentration} derived an explicit expression for the auxiliary function $g(r')$ under the assumption that $\chi_w(r)$ can be represented as a polynomial. This allows a symbolic computation of the evaporation rate $J^*$ using mathematical software. However, when strong spatial gradients in $\chi_w$ are present, polynomial approximations tend to introduce spurious oscillations in the evaporation rate. To overcome this limitation, we adopt a semi-analytical approach that is robust regardless of the smoothness of $\chi_w$. Specifically, we represent $\chi_w$ using cubic splines, which permits an analytical solution for $g(r')$, followed by a numerical integration to evaluate $J^*$. The strength of this method lies in the ability of cubic splines to approximate arbitrary continuous functions without inducing artificial oscillations. Technical details of the spline-based implementation are provided in Appendix~\ref{app:EvRate}.

\subsection{Non-dimensionalization}

The characteristic scales of the problem are those induced by evaporation. Thus the characteristic velocity is given by evaporation as $v_c=J_c^*/\rho^*\epsilon$, where $J_c^* =D_{vap}C_s(1-H_r)/R_c$. Based on this scale, we introduce the following dimensionless variables:

\begin{equation}
\begin{array}{c}
r = \dfrac{r^*}{R_c^*}, \quad z = \dfrac{z^*}{\epsilon R_c^*}, \quad \bar{u}_r = \dfrac{\epsilon \rho^* \bar{u}_r^*}{J_c^*}, \quad t = \dfrac{J_c^* t^*}{\epsilon \rho^* R^*}, \\[0.5em]
h = \dfrac{h^*}{\epsilon R^*}, \quad p = \dfrac{(p^* - p_{\mathrm{atm}}^*) \rho^* R^* \epsilon^3}{\mu_w^* J_c^*}, \quad \mu = \dfrac{\mu^*}{\mu_w^*}, \quad J = \dfrac{J^*}{J_c^*}.
\end{array}
\end{equation}
On this dimensionless time a pure water droplet is fully evaporated at $t_{ev,w}=\pi/8$ \citep{moore}. Notice that, a priori, the viscosity varies with protein concentration. Therefore, we choose the viscosity of pure water, $\mu_w^*$, as the characteristic viscosity for non-dimensionalization. Introducing this non-dimensional variables the flow problem becomes 

\begin{equation}\label{eq4:NonDimensionalModel}
\frac{\partial h}{\partial t}+\frac{1}{r} \frac{\partial}{\partial r}\left(r h \bar{u}_r\right)=-J, \quad \bar{u}_r=-\frac{h}{3\mu Ca} \frac{\partial p}{\partial r}, \quad p=-\frac{1}{r} \frac{\partial}{\partial r}\left(r \frac{\partial h}{\partial r}\right),
\end{equation}
subjected to symmetry boundary conditions at the symmetry axis

\begin{equation*}
    \frac{\partial h}{\partial r}=Q = 0  \  \mathrm{at} \ r=0, 
\end{equation*}
and no-flux and zero-height condition at the contact line

\begin{equation*}
    h=Q= 0  \  \mathrm{at} \ r=1.
\end{equation*}
with the initial condition 

\begin{equation*}
    h = 1-r^2  \  \mathrm{at} \ t=0.
\end{equation*}

Here $Ca=J_c^*\mu_w^*/\gamma^*\rho^*\epsilon^4$ is the capillary number. The Capillary number is typically small for water-based droplets evaporating under ambient conditions. In our experiments, for instance, we estimate $ Ca \sim 10^{-6} \ll 1 $, indicating that surface tension dominates over viscous forces, and the droplet interface is expected to remain close to a paraboloid of revolution. However, as we will discuss in subsequent sections, the viscosity of protein solutions can increase significantly with concentration. In such cases, the product $ \mu Ca $ becomes of order unity or larger, suggesting that viscous effects could deform the interface away from the paraboloid shape \citep{coombs2024colloidal}.

Introducing this non-dimensional variables the solute problem becomes 

\begin{equation}\label{eq4:SoluteDimensionaless}
\frac{\partial}{\partial t}\left(h \overline{w}_p\right)+\frac{1}{r} \frac{\partial}{\partial r}\left(r h \bar{u}_r \overline{w}_p\right)=\frac{1}{Pe}\frac{1}{r} \frac{\partial}{\partial r}\left(D_pr h \frac{\partial \overline{w}_p}{\partial r}\right),
\end{equation}
where we have used the diffusivity of the protein in a very dilute solution, denoted $D_{p_0}^*$, as the characteristic diffusivity for non-dimensionalization. Accordingly, the dimensionless diffusivity is defined as $ D_p = D_p^*/D_{p_0}^* $, where $D_p^*$ is the local protein diffusivity. This equation is subjected to symmetry conditions at the symmetry axis

\begin{equation*}
    \frac{\partial \overline{w}_p}{\partial r} = 0  \  \mathrm{at} \ r=0, 
\end{equation*}
and no-flux condition at the contact line

\begin{equation*}
    h \left(\bar{u}_r \overline{w}_p-\frac{1}{Pe} \frac{\partial \overline{w}_p}{\partial r}\right)=0  \;  \mathrm{at} \; r=1,
\end{equation*} 
with initial condition

\begin{equation*}
    \overline{w}_p = w_{p_0} \  \mathrm{at} \ t=0.
\end{equation*}

The Péclet number is defined as $Pe = J_c^* R_c^* / \rho^* D_{p_0}^* \epsilon$. The formation of a protein rim requires $Pe \gg 1$, which ensures that the advective mass flux is strong enough to overcome the opposing diffusive flux.  

Finally the non-dimensional evaporation rate can be computed by 

\begin{equation}
J=-\frac{1}{r} \frac{\partial}{\partial r} \int_{r}^{1} \frac{r' g(r') \, \mathrm{d}r'}{\sqrt{r'^2-r^2}}
\end{equation}
where 

\begin{equation}
g(r')=\frac{2}{\pi} \frac{\mathrm{d}}{\mathrm{d} r'} \int_0^{r'} \frac{r \left(\,\chi_w(r)-H_r\right) \, \mathrm{d}r}{\sqrt{r'^2-r^2}}.
\end{equation}

For details of the numerical scheme used to solve this dimensionless model, the interested reader is referred to Appendix~\ref{app:NumSch}.  

\subsection{Water activity and material properties}\label{sec:WatActMatProp}

Water activity is known to depend on both protein and salt mass fractions \citep{znamenskaya, Mikhailov2004}. To obtain a realistic representation of the water activity $\chi_w(\bar{w}_p, w_s)$, we utilize the framework of the Ross equation \citep{CazierWatActivityMultSystems}, which assumes non-interacting solutes. In this approach, the water activity of a multicomponent system is the product of the individual activities: $\chi_w \approx \chi_w^p(\bar{w}_p) \cdot \chi_w^s(w_s)$. For solutions near the dilute limit ($\chi_w \approx 1$), this expression can be linearized via a Taylor series expansion. By neglecting higher-order terms, the water activity is expressed as:

\begin{equation}\label{eq:watact}
    \chi_w(\bar{w}_p, w_s) = \chi_w^p(\bar{w}_p) + \chi_w^s(w_s) - 1,
\end{equation}

where $\chi_w^p(\bar{w}_p)$ and $\chi_w^s(w_s)$ are the empirical water activities in the presence of protein \citep{znamenskaya} and salt \citep{Mikhailov2004}, respectively. While this linear model is strictly valid for small deviations from ideality, it provides a decoupling of the different solute contributions. This decoupling is essential for our framework to account for the spatial segregation of mucin and salt during evaporation—a phenomenon where local variations in concentration drive changes on the evaporation flux independently. Such differences in solute concentration would not be captured by global empirical measurements of multicomponent systems, such as those reported by \cite{merhi2022assessing} for real saliva, in which water activity is described as a function of the total solute concentration.

As discussed in Section~\ref{sec:CompExp}, comparison with our experimental data indicates improved agreement when a quasi-ideal form is used for $\chi_w^p(\bar{w}_p)$. Throughout this work, however, the functional form we used is the one proposed by \cite{Salmon}, fitted with the experimental data of \cite{znamenskaya} (see appendix~\ref{app:FitExperimentalData} for further details on the implementation). This is leading to $A_{wB}$=3.39 and $B=3$ in the functional form proposed by \cite{Salmon}:
    \begin{equation*}
    \chi_w^p = (1 -\bar{w}_p)e^{\bar{w}_p + \xi \bar{w}_p^2},
    \end{equation*} where $\xi = A_{wB} - B(1 - \bar{w}_p)^{0.09}$.

This values of the water activity associated with protein, $\chi_w^p$, are used except in Section~\ref{sec:CompExp}. For a detailed discussion of alternative water activity models, the reader is referred to Appendix~\ref{app:AwB}. 

We do not include an explicit transport equation for salt in our formulation. This simplification is justified by the high salt diffusivity, $D_s^* = 10^{-9}\,\text{m}^2/\text{s}$, which yields Péclet numbers on the order of $Pe_s \sim 1$. Thus, we assume that the salt remains well mixed throughout the droplet at all times \citep{seyfert2022stability, martinezpuig2025}. Given the initial salt mass fraction $w_{s_0}$ and the initial droplet volume $V_0$, the total salt mass is $M_s = w_{s_0} V_0$. The droplet volume at any time is  $V(t) = 2\pi \int_0^1 r \, h(r,t) \, dr,$ which allows us to compute the instantaneous salt mass fraction as $w_s(t) = M_s / V(t)$, provided that the salt mass is conserved due to its non volatility.

The difference in diffusivity between salt and protein is crucial in the formation of deposition patterns. The relatively low diffusivity of proteins enables the development of strong spatial concentration gradients near the contact line, which in turn makes the formation of a protein-rich ring possible. In contrast, the high diffusivity of salt ensures that it remains well mixed throughout the droplet, preventing the buildup of localized concentration gradients. This distinction is clearly reflected in the experimental salt deposition patterns (see figure~\ref{fig:ResExp}). Unlike the protein, which accumulates locally to form a concentrated ring near the contact line, salt crystals are observed throughout the entire droplet footprint. For a more thorough discussion and modeling of salt transport, we refer the interested reader to \cite{martinezpuig2025}.

\begin{figure}
    \centering
    \includegraphics[width=\linewidth]{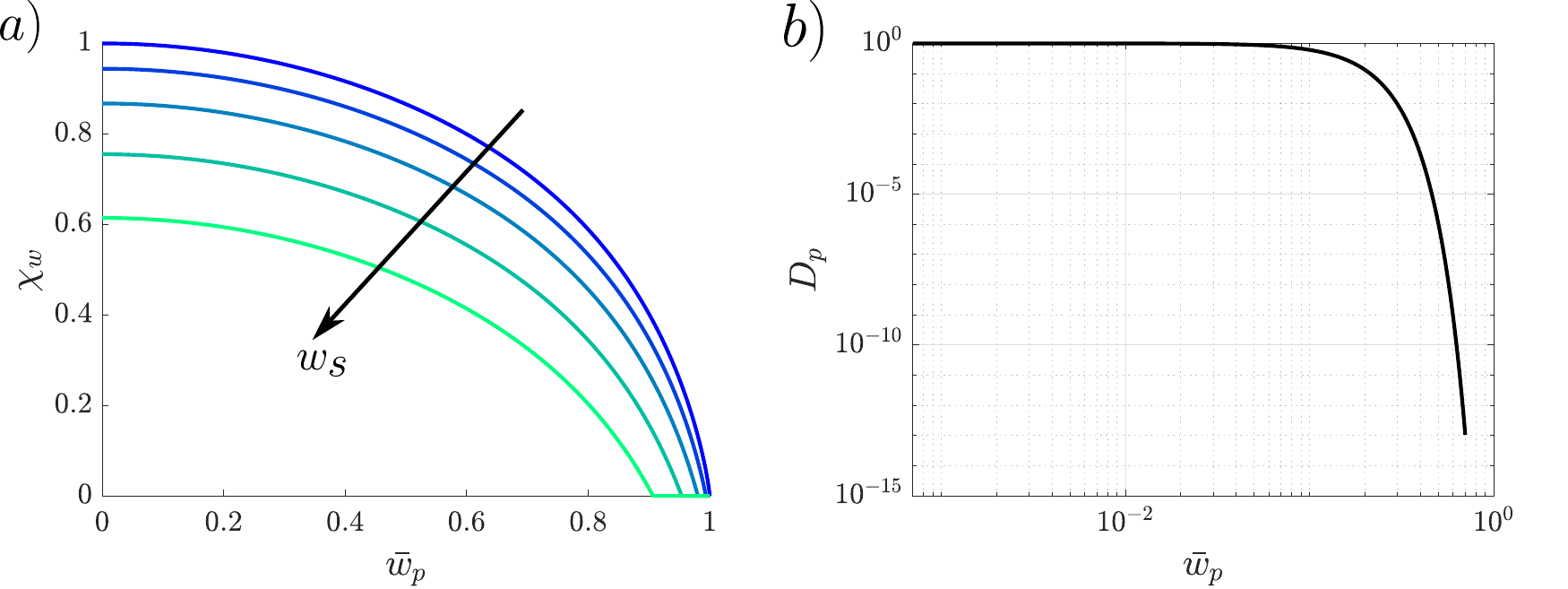}
    \caption{Material and thermodynamical properties used in the model. (a) Water activity as a function of protein mass fraction for different salt mass fractions $w_s = 0, 0.09, 0.18, 0.26, 0.35$. Increasing salt concentration shifts the water activity curves downward. (b) Evolution of protein diffusivity as a function of protein concentration, display on a logarithmic scale.}
    \label{fig:Matprop}
\end{figure}

From the perspective of water activity, this has an important implication: the effect of salt on water activity is the same throughout the interface as the salt concentration is homogeneous throughout the droplet \cite{seyfert2022stability}. In contrast, the contribution of protein to water activity is local, affecting only regions where the protein mass fraction becomes sufficiently high. This localized influence can lead to spatially heterogeneous evaporation rates, which drive the advective transport necessary for ring formation. In figure~\ref{fig:Matprop} a) we show the dependence of water activity on both protein and salt mass fractions. 

We use a diffusivity similar to that proposed by \cite{Salmon} for complex fluids. In our case, we use 

\begin{equation}
\log\left(D_p^*(\bar{w}_p)/(1\,\mathrm{m}^2/\mathrm{s})\right)=a_4 \bar{w}_p^4+a_3 \bar{w}_p^3+a_2 \bar{w}_p^2+a_1 \bar{w}_p+a_0,
\end{equation}

where $\left[a_4 ; a_3 ; a_2 ; a_1 ; a_0\right] = \left[-14.65; 5;-22.97; 0.12;-11\right]$ (see figure~\ref{fig:Matprop} b). Compared to the constants used by \citet{Salmon}, we have chosen $a_0 = -11$, as suggested by the data of \citet{BansilViscosityMucin}, and adjusted the other constants to ensure a decay of the diffusivity coefficient for mass fractions approaching unity. We note that, once the Péclet number is small, the diffusivity coefficient is primarily relevant for capturing the freezing of the protein rim at high relative humidity (see Section~\ref{sec:SaltNonConsDiff} for a detailed discussion on the role of the diffusivity coefficient).

The density of water, salt and mucin solution has been reported by \cite{mikos1989}. Although the density varies with composition, the relative variations remain small, and since $\Delta \rho / \rho_0 \ll 1$, we adopt the approximation of constant density in our model.The viscosity is taken to be that of water. Within the lubrication approximation, viscosity only plays a role when $O(\mathrm{Ca}) \sim 1$ or $O(\mathrm{Ca}) \gg 1$. In the regime relevant here, the flow rate is determined by the condition of minimal surface area in the limit $O(\mathrm{Ca}) \ll 1$. We expect this limit to be valid for most of the evaporation process, since initially $\mathrm{Ca} \sim 10^{-6}$. Only at very late times does the viscosity increase significantly due to protein glass transition. The possible inclusion of a viscosity dependence on protein concentration is discussed in section~\ref{sec:limitations}. Numerical limitations prevent us from incorporating this effect into the model

\subsubsection{Real saliva properties}

In order to verify that the same physical framework also applies to real saliva, we compare our results with those obtained using the water activity and diffusivity data reported by \cite{merhi2022assessing}. For this purpose, in Section~\ref{sec:salivaResults}, we adopt the functional form for water activity proposed in \cite{merhi2022assessing}. For the diffusivity we fitted the diffusivity data reported in \cite{merhi2022assessing} using the same polynomial functional form introduced by \cite{Salmon} (see Appendix~\ref{app:FitExperimentalData} for details of the fitting procedure). The resulting fitted coefficients are $[a_4,a_3,a_2,a_1,a_0]=[-9.34,\,10.89,\,-3.63,\,0.838,\,-10.58]$. Since the data reported by \cite{merhi2022assessing} are expressed as a function of the total solute concentration, we use $w_{p_0}=12\times10^{-3}$ in our model.

\section{Model results and comparison with experiments}\label{sec:Results}

\subsection{Formation mechanism of the protein ring}

To describe the dynamics of a nascent coffee ring, \cite{moore} presented a model similar in structure to the one developed here, but in which the hydrodynamics and solute transport are decoupled. In the asymptotic limits $\mathrm{Ca} \ll 1$ and $\mathrm{Pe} \gg 1$, they showed that the protein ring sharpens and ultimately converges to a singular line as evaporation progresses. This naturally raises the question: how can a finite ring width emerge without imposing a constraint that would be artificial for polymeric or protein solutes, such as a maximum packing fraction, as proposed by constant-activity models?

\begin{figure}[ht!]
    \centering
    \includegraphics[width=\textwidth]{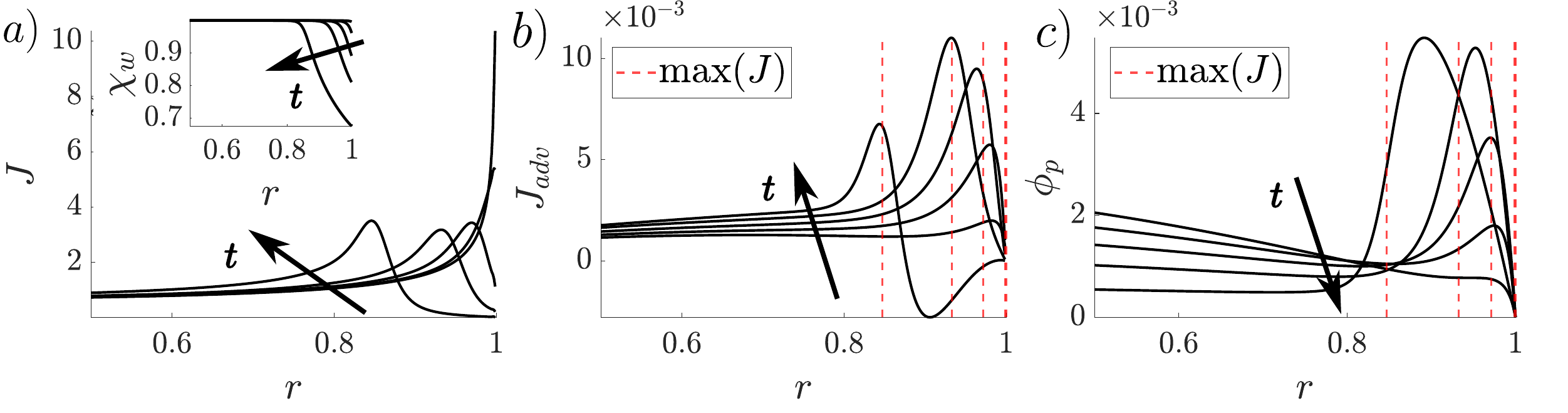}
    \caption{Mechanism of ring formation during droplet evaporation. (a) Temporal evolution of the evaporation rate, with the inset showing the corresponding evolution of water activity. (b) Temporal evolution of the advective flux and the maximum evaporation rate. (c) Temporal evolution of the height-integrated protein fraction together with the maximum evaporation rate. Data are shown at non-dimensional times
$t = t_{ev,w}\times(0.1, 0.3, 0.5, 0.7, 0.9)$, with parameters $Ca=10^{-4}$, $Pe = 13.2$, $w_{p_0}=3 \cdot10^{-3}$ and $H_r = 0.7$.}
    \label{fig:MecRing}
\end{figure}

In our model, the growth of the ring width is a natural consequence of the coupling between solute transport and hydrodynamics through an activity-dependent evaporation rate. In order to understand the proposed minimal mechanism behind the formation and widening of the ring, let us assume for now that the physical properties are constant and that there is no salt in the mixture, so we let $\mu=D_p=1$ and $w_{s_0}=0$. In figure \ref{fig:MecRing} we show the temporal evolution of the evaporation rate $J$, the water activity $\chi_w$, the advective flux $J_\mathrm{adv}=\bar{u}_r h w_p$, and the height-integrated protein fraction $\phi_p$, for a simulation with $Ca=10^{-4}$, $Pe = 13.2$, $w_{p_0}=3 \times 10^{-3}$, and $H_r = 0.7$.

At early stages, when the water activity $\chi_w \approx 1$ everywhere, the evaporation rate presents a sharp maximum near the contact line, driving a strong advective flux of protein towards the edge which, in turn, leads to the local accumulation of solute. If the evaporation rate does not depend on the solute concentration, then it diverges to infinity at the contact line at all times. However, this situation is modified when we let the water activity be a decreasing function of the solute concentration. As a result of the large evaporation rate near the contact line, water is quickly depleted there, which diminishes the evaporation rate (figure \ref{fig:MecRing}a) and regularizes the singularity. Also, the maximum evaporation rate shifts radially inwards, away from the contact line. It is worth mentioning here that, as has been previously discussed for the formation of the coffee ring in particle-laden droplets, the presence of a singularity in the evaporation rate is not essential to create a coffee ring \citep{BoulogneSing, WilsonSing}, as we also show here. 

In the limit of small capillary number, the dominant surface tension forces impose a spherical-cap interface at all times. Mass conservation then uniquely determines the flow rate $Q$ for $\partial h / \partial t$ to be compatible with a spherical-cap interface. In particular, it dictates that the maximum flow rate is going to be close to the location where the evaporation rate peaks.

In fact, in the case $\mu = 1$, the water flux demanded by the evaporation peak can be as strong as to reverse the flow in the region between the maximum evaporation rate and the contact line, which corresponds to the protein rim, figure \ref{fig:MecRing}b. Although this effect, which we dub \textit{protein rim squeezing}, is regularized if we let the viscosity grow large enough with the protein concentration (see section~\ref{sec:limitations} for a detailed discussion, and also \cite{coombs2024colloidal}).
In any case, even if the flow rate never becomes negative, the strong advective flux directed to the evaporation rate peak transports protein mostly towards the inner region of the ring, promoting its widening. In fact, in figure \ref{fig:MecRing}c we see how the protein ring extends from the contact line to the location of maximum of the evaporation rate.

\begin{figure}[ht!]
    \centering
    \includegraphics[width=\textwidth]{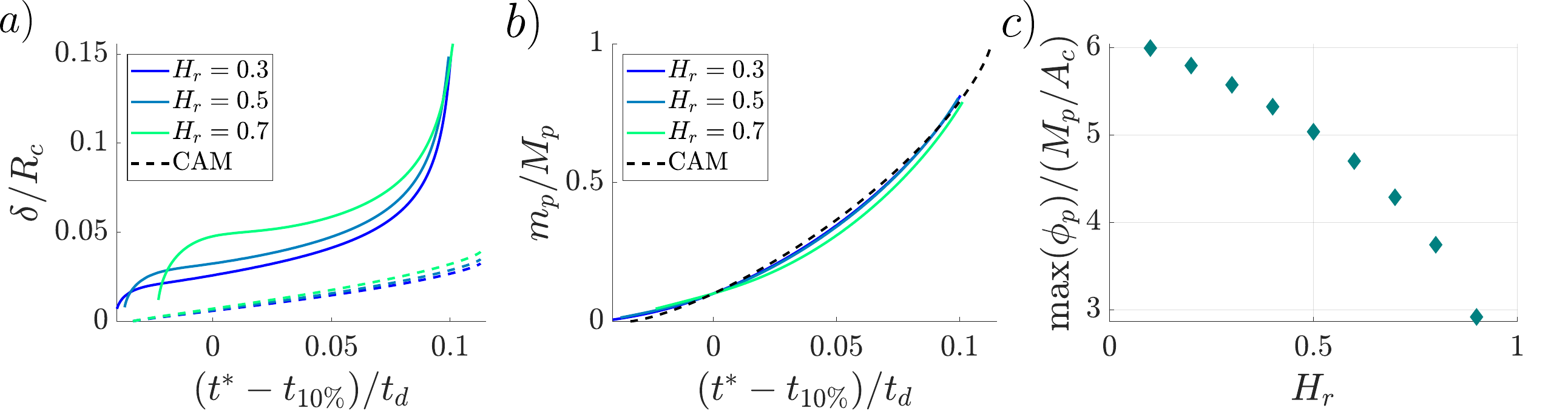}
    \caption{Evolution of the ring morphology during droplet evaporation. (a) Temporal evolution of the ring width. (b) Temporal evolution of the protein mass in the ring, normalized by the total protein mass in the droplet. (c) Maximum height-integrated protein mass fraction, normalized by the total protein mass in the droplet divided by the contact area $A_c$, as a function of relative humidity. The parameters used are $Ca = 10^{-4}$, $Pe = 43.3(1 - H_r)$, initial protein mass fraction $w_{p_0} = 3 \cdot 10^{-3}$, and relative humidity values $H_r = 0.3, 0.5, 0.7$. Simulations are stopped at $t=t_{ev,w}$.}
    \label{fig:WidthRing}
\end{figure}

In figures~\ref{fig:WidthRing}a and \ref{fig:WidthRing}b we show the model results, which portrait the time evolution of the rim width and mass for three different values of the relative humidity. We use the same parameters as in figure~\ref{fig:MecRing}, considering relative humidity values of $H_r = 0.3,\ 0.5,\ \text{and}\ 0.7$. Additionally, we present the evolution of the final maximum of the height-integrated protein fraction $\phi_p$ as a function of the relative humidity. Despite its simplicity, our minimal model —where salt is neglected and material properties are constant— qualitatively reproduces the most salient features observed in experiments: the dependence of the final rim width with the relative humidity, the insensitivity of rim mass evolution to the relative humidity, and the decreasing trend of the maximum height-integrated protein fraction with the relative humidity.

We can explain why, containing nearly the same mass, the rim width grows monotonically with the humidity. The relative humidity appears in the problem through the difference $\chi_w - H_r$ (Equations \eqref{eq:J_vs_integral_g}-\eqref{eq:g_vs_integral_chiw}). As this factor approaches zero near the contact line (becoming instantaneously zero there at $t=0^+$), the peak of maximum evaporation rate smoothens up, and the location of maximum evaporation shifts inward. This transition occurs earlier the higher the relative humidity, because less protein needs to accumulate near the contact line to satisfy the condition $\chi_w(w_p) - H_r \ll 1$. As a result, the ring width increases more rapidly the higher the humidity. Then, since the ring mass is rather insensitive to the humidity, the maximum height-integrated protein concentration must therefore decrease.

To compare the measured ring width with the theoretical prediction of constant-activity models (CAM), we will use the work of \citet{Popov}.  We need to impose a maximum protein concentration, equivalent to a maximum particle packing fraction. Nevertheless, CAM predict ring widths more than three times smaller than those obtained with our model (figure \ref{fig:WidthRing}a). This discrepancy can be attributed to two main factors. First, the protein concentration within the rim is not homogeneous. Second, the maximum protein concentration in the ring evolves over time rather than remaining constant. Since the protein concentration in our model remains, for most of the time and across most of the ring, below the maximum concentration reached during the entire evaporation process, the rim must be thicker than in Popov’s description.

The ring mass, however, compares reasonably well with the theoretical prediction of \citet{Popov}, originally derived by \citet{deegan2000contact}. In the absence of diffusion, the ring mass is determined by the advective flux driven by evaporation, implying that it should correlate with the total mass of water lost from the droplet. Although the maximum of the evaporation rate is shifted toward the inner region of the ring---leading to a thickening of the deposit---the overall ring mass remains close to Popov's prediction, provided that the total evaporated mass is similar. The same reasoning also explains why the evolution of protein mass in the ring shows little dependence on relative humidity: since the volume evolution is nearly unaffected by the humidity when represented against the non-dimensional time, the total advective flux required to replenish water losses near the contact line is also similar.


The influence of the humidity in the protein ring evolution depends on the functional form of the water activity $\chi_w^p(\bar{w}_p)$. For example, if an ideal binary mixture is assumed, such that $\chi_w^p(\bar{w}_p) \sim 1 - w_p$~\citep{merhi2022assessing}, then the relationship between humidity and rim formation becomes nearly linear. However, this is not the case in more complex solutions, which depart from ideal behavior and account for non-equilibrium effects. In our model, based on a non-ideal $\chi_w^p(\bar{w}_p)$, the difference in behavior between medium and high relative humidity is more pronounced than that between low and medium. This is because the critical concentration $w_{p_c}$ satisfying $\chi_w(w_{p_c}) = H_r$ is similar for low and medium humidity, but differs significantly for high humidity (see appendix \ref{app:AwB} and subsection \ref{sec:SphericalIdeal} for a detailed discussion on different water activities).

Although our minimal model qualitatively captures the main trends observed in the experiments, additional complexity is required to fully explain the differences between figure~\ref{fig:WidthRing} and figure~\ref{fig:ResExp}. In the following subsection, we analyze the combined influence of salt and non-constant diffusivity.

\subsection{Role of salt and non-constant diffusivity.}\label{sec:SaltNonConsDiff}

In our experiments, we observe that droplets evaporated at high relative humidity exhibit a freezing of the ring structure at late times. This is evidenced by the plateaus in both the width and mass evolution of the ring for $H_r \approx 70\%$, (see figures~\ref{fig:ResExp} a and b). We also find that, at high humidity, some droplets remain in a liquid state for an extended period, until we abruptly decrease the humidity to that of the ambient at the end of the experiment. A similar behavior at high humidity has been previously observed in droplets of the same composition evaporating on superhydrophobic substrates \citep{seyfert2022stability}.

This behavior can be explained by the presence of salt. As evaporation proceeds, the salt concentration within the droplet increases until a critical concentration is reached such that $\chi_w(w_s) = H_r$, effectively halting evaporation \citep{seyfert2022stability}. Unlike protein-driven evaporation arrest, which is governed by a local condition $\chi_w(w_p) \to H_r$ near the contact line, salt-induced evaporation arrest is a global phenomenon. This is because the high diffusivity of salt ensures it remains well mixed throughout the droplet, thereby reducing the water activity uniformly and halting evaporation across the entire interface.

\begin{figure}[ht!]
    \centering
    \includegraphics[width=\textwidth]{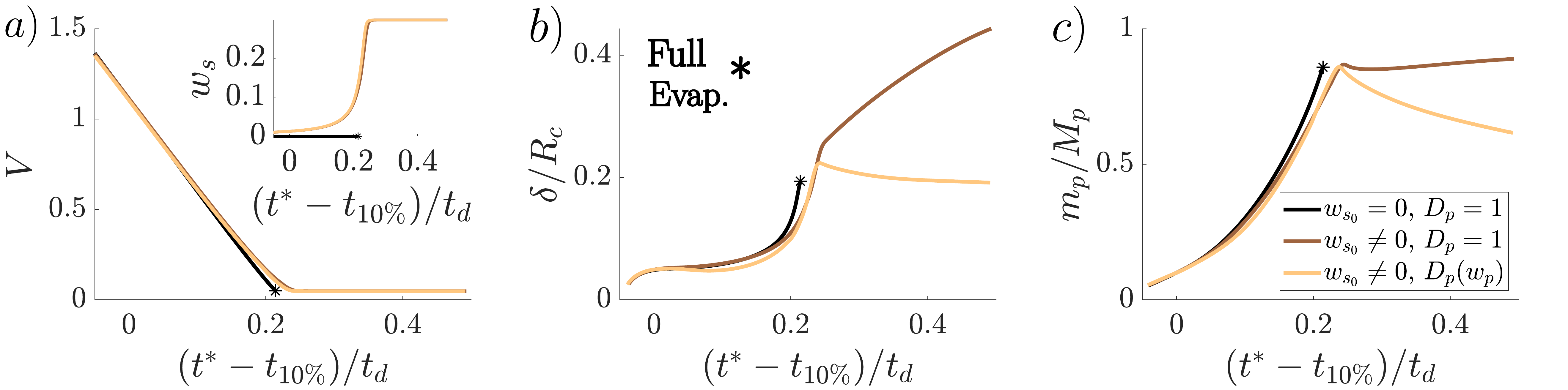}
    \caption{Impact of salt presence and non-constant protein diffusivity on the morphology of the protein ring at high relative humidity. (a) Temporal evolution of the droplet volume, with the inset showing the corresponding salt mass fraction. (b) Evolution of the protein ring width. (c) Protein mass in the ring normalized by the total protein mass in the droplet. Simulations use parameters $Ca = 10^{-4}$, $Pe = 13.2$, initial protein mass fraction $w_{p_0} = 3 \times 10^{-3}$, relative humidity $H_r = 0.7$. Results compare different initial salt mass fractions $w_{s_0} = 0, 9\cdot10^{-3}
    $ and constant protein diffusivity $D_p = 1$ and concentration-dependent diffusivity $D(w_p)$. Simulations are extended until $t=2t_{ev,w}$ when $w_{s_0} = 9\cdot10^{-3}$ to see the stabilization of the volume.}
    \label{fig:NablaDp}
\end{figure}

However, we did not observe the same behavior for all experiments performed at the same conditions of high humidity. Some droplets evaporated completely within the experimental time, while others remained in a liquid state as long as the experiment lasted. This variability arises because the condition $H_r = \chi_w(w_{s_c})$ corresponds to a salt concentration $w_{s_c}$ above the saturation concentration. Although crystallization is known to occur in evaporating salt-containing droplets at supersaturation \citep{DesMetastability, SHAOCrystDroplets}, the system may remain in a metastable liquid phase once salt concentration is above saturation. In this state, even small fluctuations in ambient relative humidity can trigger efflorescence (i.e., salt crystallization). This delicate balance explains why some droplets at $H_r = 0.7$ crystallize while others remain liquid until the end of the experiment.

To assess the influence of salt, we incorporate into our model an initial salt concentration of $w_{s_0} = 9 \cdot 10^{-3}$, as in our experiments. We observe that, under these conditions, droplets evaporating at $H_r = 0.7$ remain stable, but only after the salt mass fraction reaches approximately $0.3$ (see figure~\ref{fig:NablaDp}a), a value higher than the saturation mass fraction $0.265$. The inclusion of salt in the model also explains the observed plateau in the final stages of the ring mass evolution (see figure~\ref{fig:NablaDp}c). Once evaporation halts due to the condition $\chi_w(w_s) = H_r$, the advective flux that transports solutes toward the contact line disappears. As a result, the solute mass accumulated in the ring remains nearly constant. Interestingly, despite the halt in evaporation, we observe that the width of the ring continues to increase. This occurs because, in the absence of advective transport, a residual diffusive flux persists, gradually spreading the protein-rich region inward into the droplet. Eventually, this leads to a broadened ring structure that can extend over half of the droplet. However, this behavior is inconsistent with our experimental observations. This discrepancy highlights the limitations of the assumption of constant diffusivity. In reality, protein diffusion is strongly concentration-dependent, especially for concentrated solutions \citep{RogerK_ContrWatEvp,Salmon, merhi2022assessing}, and a more accurate model must account for a non-constant diffusivity to prevent such unrealistic ring broadening.

This result is particularly interesting because, unlike the case of spherical droplets, the formation of a stable polymeric or protein rim in sessile droplets does not necessarily require a concentration-dependent diffusivity for low and medium ambient relative humidity. In spherical droplets, the entire interface becomes uniformly covered by the dense protein shell, effectively halting the apparent advection caused by the receding interface. This shell formation is observed across all values of relative humidity $H_r$. However, if the diffusivity remains constant, the shell can diffuse inward over a time scale, $t_d = R_c^2 / D_p \sim 10^3$ \SI{}{\second} for a droplet of radius $R_c =$ \SI{100}{\micro\meter} and protein diffusivity $D_p = 10^{-11}$ \SI{}{\meter\squared\per\second}. This time scale is comparable to its evaporation time.

In contrast, for sessile droplets, the ring formation is not governed by the apparent advection from interface recession but rather by the advective flux required to maintain a pinned contact line and a spherical-cap shape. As a result, the protein ring forms in the vicinity of the contact line, leaving a large portion of the droplet interface free of solute and still contributing to evaporation. Consequently, for low and medium values of $H_r$, the dependence of the diffusivity on the protein concentration is not critical. As long as the Péclet number satisfies $Pe \gg 1$, the solute accumulates at the rim and salt crystallizes, as evaporation proceeds, before the ring has time to diffuse inward.

However, at high relative humidity, evaporation is globally arrested not due to protein concentration but due to salt concentration, which modifies the water activity such that $\chi_w(w_s) = H_r$ is reached almost uniformly throughout the droplet. Under these conditions, diffusivity plays a crucial role in maintaining a stable ring structure. The diffusive flux must be significantly reduced so that the associated diffusive timescale becomes much longer than the experimental timescale. This is indeed consistent with our experimental findings.  When we incorporate a non-constant diffusivity $D_p(w_p)$, which decreases sharply at high protein concentrations (see Section \ref{sec:WatActMatProp}), we observe that the ring width remains stable over long times (see figure~\ref{fig:NablaDp} b). This highlights the importance of accounting for concentration-dependent transport properties in modeling the late-stage dynamics of evaporating droplets under high humidity.

\subsection{Spherical vs non-spherical droplets. Ideal vs non-ideal water activity}\label{sec:SphericalIdeal}

\begin{figure}[ht!]
    \centering
    \includegraphics[width=\textwidth]{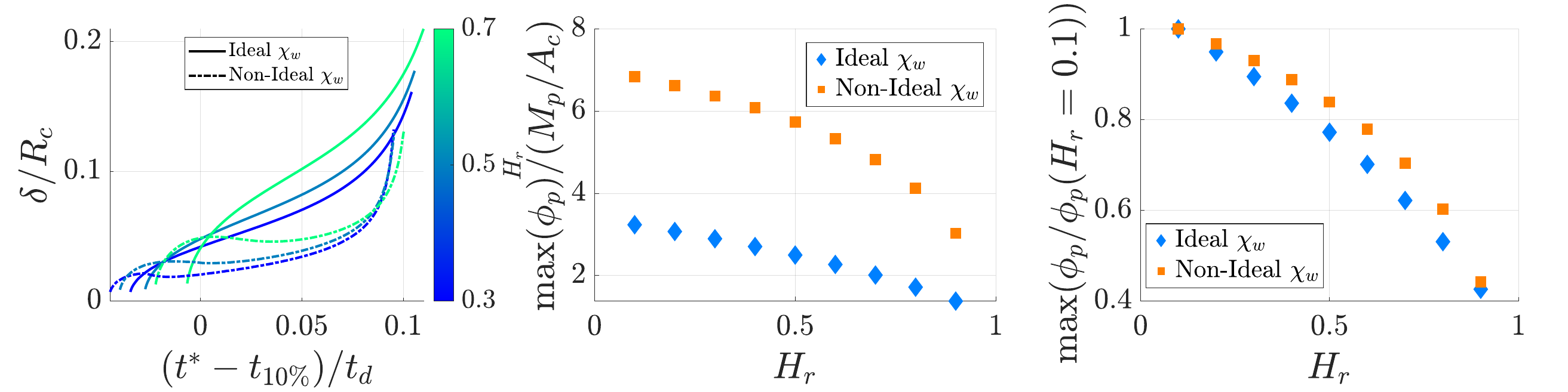}
    \caption{Comparison of the ring parameters when using an ideal ($\chi_w^p = 1-w_p$) and a non-ideal (see section~\ref{sec:WatActMatProp}) water activity. (a) Temporal evolution of the ring width. (b) Maximum height-integrated protein mass fraction, normalized by the total protein mass in the droplet divided by the contact area $A_c$, as a function of relative humidity. (c) Relative changes in the maximum height-integrated protein mass fraction, as a function of relative humidity. The parameters used are $Ca = 10^{-4}$, $Pe = 43.3(1 - H_r)$, initial protein mass fraction $w_{p_0} = 3 \cdot 10^{-3}$, and relative humidity values $H_r = 0.3, 0.5, 0.7$. Simulations are stopped at $t=t_{ev,w}$.}
    \label{fig:Ideal_vs_NonIdeal}
\end{figure}

The sensitivity of the final deposition patterns to relative humidity is closely linked to the assumed dependence of the water activity on solute concentration for spherical droplets \citep{merhi2022assessing,HautRespDropDyn}. To clarify wether this is the case in sessile droplets, we compare the cases of an ideal water activity, $\chi_w^p = 1 - w_p$, and a non-ideal water activity, as detailed in section~\ref{sec:WatActMatProp}. For sessile droplets, we find that the rim morphology remains sensitive to relative humidity (see figure~\ref{fig:Ideal_vs_NonIdeal}). The maximum protein concentration decreases as $H_r$ increases, although this decrease is more pronounced when an ideal water activity is assumed (see figure~\ref{fig:Ideal_vs_NonIdeal}c). This shows that, while a non-ideal water activity quantitatively affects the rim characteristics, the qualitative behavior observed in the experiments can be explained by our model without having to assume a strong non-ideality in the water activity.

This behavior contrasts sharply with that of spherical droplets, as reported in previous experimental and theoretical studies \citep{merhi2022assessing, HautRespDropDyn}. For spherical droplets, the symmetry of the problem reduces the evaporation dynamics to a one-dimensional description, in which the evaporation flux depends linearly on the difference between the (uniform) water activity and the ambient relative humidity, $J \sim \chi_w - H_r$. This symmetric, and thus uniform, dependence explains why, when a non-ideal water activity is considered, the final solute concentration distribution becomes largely independent of relative humidity below a threshold value of approximately $70\%$: the solute concentration required to satisfy the condition $\chi_w - H_r = 0$ is nearly identical over this range of humidities.

By contrast, in sessile droplets the evaporation rate depends on the local value of the difference $\left(\chi_w - H_r\right)$ (see equations \ref{eq:J_vs_integral_g} and \ref{eq:g_vs_integral_chiw}). Moreover, solute-induced arrest of evaporation is a local phenomenon in sessile droplets, occurring preferentially near the contact line where protein accumulates, whereas in spherical droplets the influence of solutes on evaporation is global, acting uniformly over the entire interface. These combined effects explain why, in sessile droplets, both the maximum protein concentration and the width of the protein rim remain sensitive to relative humidity over the entire range of humidity investigated, in contrast to the behavior observed for spherical droplets of similar compositions.

\subsection{Comparison with experiments}\label{sec:CompExp}

Direct comparisons between experimental measurements and hydrodynamic–solute transport models for the spatio-temporal evolution of solute concentration in evaporating multicomponent droplets remain scarce. While some studies have compared experiments and simulations in multicomponent systems, these have primarily focused on predicting particle distributions rather than resolving solvent or solute concentration fields \citep{RajuMarRingWatGlyc}. This limitation stems from the experimental difficulty of measuring concentration profiles in evaporating droplets, as well as from the fact that existing models—largely developed for particle-laden droplets—do not capture key features of ring formation in multicomponent solutions. In this section, we compare our model with the experiments presented in Section~\ref{sec:Experiments}, and discuss its limitations. The model presented in Section~\ref{sec:RedModel} contains no free parameters, and whenever possible it relies on material and thermodynamic properties from experimental studies reported in the literature (see Section~\ref{sec:WatActMatProp}).

\begin{figure}[ht!]
    \centering
    \includegraphics[width=\textwidth]{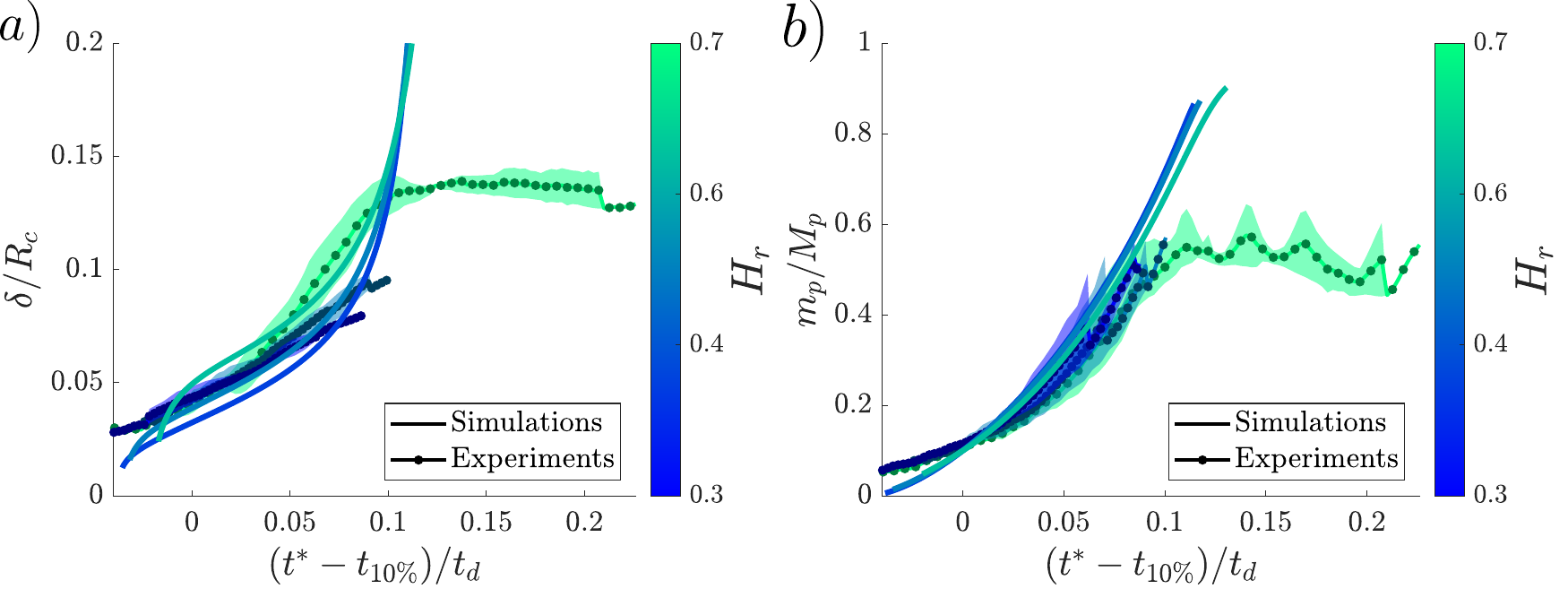}
    \caption{Comparison of experimental results with numerical predictions of our model. The solid line represents the mean of experiments performed at comparable relative humidity, while the shaded region spans the range between the minimum and maximum experimental values. (a) Temporal evolution of the ring width. (b) Temporal evolution of the protein mass in the ring, normalized by the total protein mass in the droplet. The parameters used are $Ca = 10^{-4}$, $Pe = 43.3(1 - H_r)$, relative humidity values $H_r = 0.3, 0.5, 0.7$, concentration-dependent diffusivity $D(w_p)$, initial protein mass fraction $w_{p_0} = 3 \cdot 10^{-3}$, and initial salt mass fraction $w_{s_0}=9\cdot10^{-3}$. Simulations are stopped when $w_s=0.25$.}
    \label{fig:Exp_vs_Sim_DeltaMass}
\end{figure}

To enable comparison with experiments, we include salt in our model and account for a non-constant diffusivity. For the dependence of water activity on protein concentration, $\chi_w^p(\overline{w}_p)$, we initially employed the interpolation described in Section~\ref{sec:WatActMatProp}. However, comparison with experiments shows that assuming a water activity closer to the ideal case provides much better agreement. Specifically, we borrow the functional form proposed by \citet{Salmon},  
\begin{equation}
\chi_w^p = (1 - \overline{w}_p)\,e^{\overline{w}_p + \xi \overline{w}_p^2}, \quad \text{with} \quad \xi = 0.5 - 3(1 - \overline{w}_p)^{0.09}.
\end{equation}  
The discrepancy may arise because the presence of salt modifies the influence of protein on water activity compared to the salt-free water–mucin measurements of \citet{znamenskaya}. Additional contributing factors could include differences in the chemical treatment of our mucin or variations in temperature. Therefore, throughout this section we adopt the quasi-ideal water activity model. A discussion of the impact of different water activity formulations on protein ring growth is provided in appendix~\ref{app:AwB} and subsection~\ref{sec:SphericalIdeal}.  

\begin{figure}[ht!]
    \centering
    \includegraphics[width=\textwidth]{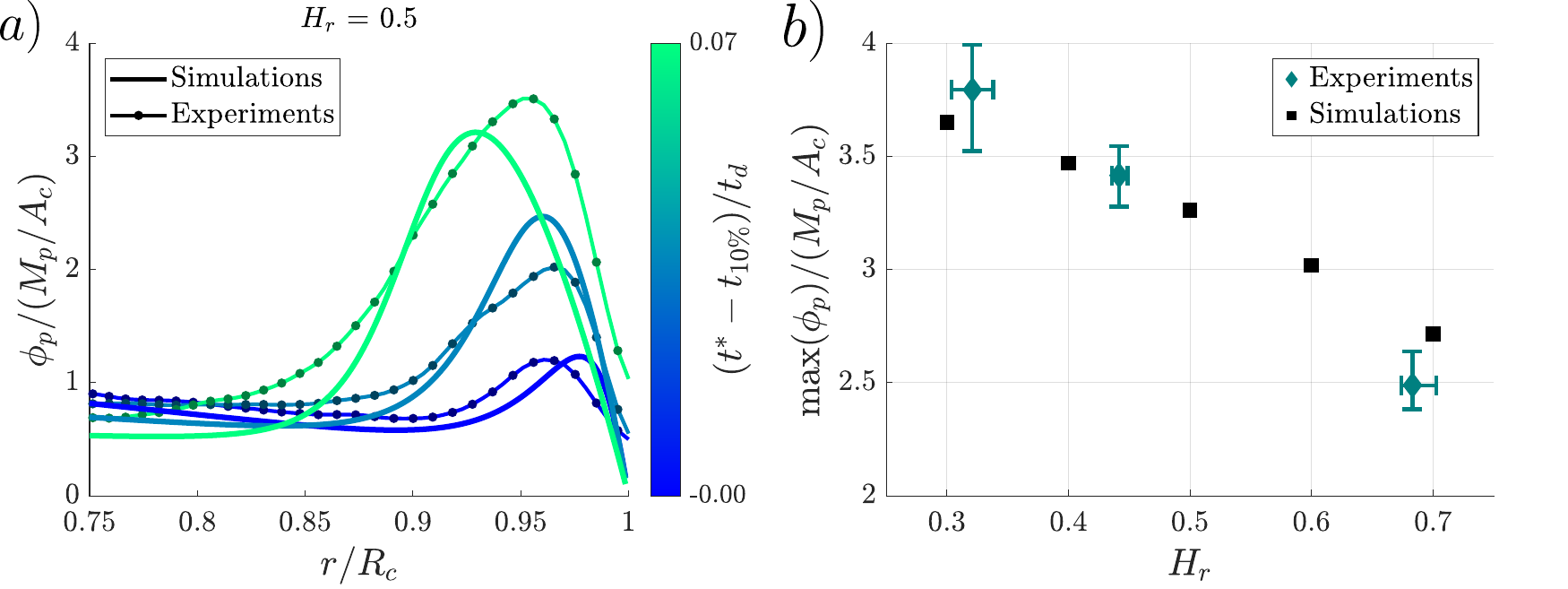}
    \caption{Comparison of experimental results with numerical predictions of our model. (a) Temporal evolution of the height-integrated protein mass fraction, normalized by the total protein mass in the droplet divided by the contact area (b) Maximum height-integrated protein mass fraction, normalized by the total protein mass in the droplet divided by the contact area, as a function of relative humidity.  The parameters used are $Ca = 10^{-4}$, $Pe = 43.3(1 - H_r)$, relative humidity values $H_r = 0.3, 0.4, 0.5, 0.6, 0.7$, concentration-dependent diffusivity $D(w_p)$, initial protein mass fraction $w_{p_0} = 3 \cdot 10^{-3}$, and initial salt mass fraction $w_{s_0}=9\cdot10^{-3}$. Simulations are stopped when $w_s=0.25$.}
    \label{fig:Exp_vs_Sim_PhiMaxPhi}
\end{figure}

Our model reproduces well the evolution of the protein ring mass (see figure~\ref{fig:Exp_vs_Sim_PhiMaxPhi}b). The evolution of the ring width is also qualitatively reproduced. The model captures the dependence of ring width on relative humidity. Interestingly, at early times it reproduces not only the offset of the curves but also the different slopes observed in the experiments for varying humidity. However, discrepancies arise later in time, but before crystallization: simulations predict a rapid rim broadening up to the point of salt saturation, whereas experiments show a slower growth. As we discuss in the following section, this mismatch likely results from the simplified rheology assumed in our model, specifically the use of a constant viscosity.  

Figure~\ref{fig:Exp_vs_Sim_PhiMaxPhi}a compares the evolution of the height-integrated protein fraction $\phi_p$ in a representative medium-humidity experiment with the corresponding simulation. The agreement is good until late times, where the \textit{squeezing effect} starts to be present. Both experiments and simulations also show that the maximum in $\phi_p$ is shifted inward from the droplet edge. However, this effect is significantly weaker in experiments than in simulations.  

Our model also captures the observed decrease in maximum height-integrated protein fraction $\phi_p$ with increasing relative humidity (figure~\ref{fig:Exp_vs_Sim_PhiMaxPhi}b). This trend cannot be explained by constant-activity models. Finally, the larger experimental variability at low humidity compared to high humidity arises from the limited temporal resolution of our imaging (one frame every five seconds). This uncertainty in identifying the last frame before crystallization introduces dispersion in cases where crystallization occurs while the ring mass is still increasing.  

We conclude that our model qualitatively explains the main features of protein rim formation: (i) rim broadening that depends on relative humidity, (ii) similar ring mass evolution across different humidities, (iii) a decrease in the maximum height-integrated protein fraction with increasing humidity. Among these phenomena, constant-activity models account only for the similar evolution of ring mass across humidities. This highlights the need to consider the detailed geometry of the protein rim in order to rationalize experimental findings. Relying solely on the ring mass evolution can be misleading, as it may suggest that growth is a consequence of a maximum protein concentration, whereas in reality rim formation results from the coupling between hydrodynamics and solute transport through the evaporation.  

However, the quantitative comparison between experiments and our model reveals a key discrepancy: the ring width is overestimated at late times. This highlights the limitations of our current model. We expect that incorporating more realistic material properties will improve the quantitative agreement with experiments.

\subsubsection{Model limitations}\label{sec:limitations}

The rheology of mucin-based solutions is an active area of research \citep{RulffMucinViscosity}. It is well established that the rheological behavior of mucin solutions depends strongly on how they are obtained. Different types of mucin exhibit distinct rheological properties, and chemical treatments can further modify their behavior \cite{GeorgiadesViscosityMucin}. Consequently, the rheological response of our mucin-based solution is not universal and cannot be directly applied to other systems. In some cases, mucin solutions behave as shear-thinning non-Newtonian fluids, while at higher protein concentrations they may form gels \citep{WaighViscosityMucin, GeorgiadesViscosityMucin}. In our model, we deliberately avoid this complexity, as our aim is to provide a minimal mechanism to explain the formation of a ``coffee-ring'' in multicomponent droplets. Nevertheless, one robust feature across rheological studies is that viscosity increases with mucin concentration, and once a critical concentration is reached the solution transitions to a gel-like state with very high viscosity \citep{BansilViscosityMucin,WaighViscosityMucin, GeorgiadesViscosityMucin}. Incorporating this effect could suppress the \textit{squeezing effect} responsible for the rapid rim broadening observed in our simulations prior to crystallization.

In our current model, viscosity is assumed constant. This implies that throughout evaporation the droplet remains in the low-capillary-number regime, where capillary forces dominate and the droplet shape can be approximated as a spherical cap. Protein accumulation near the contact line eventually suppresses evaporation locally, such that $J \sim 0$. However, the temporal derivative of the height remains nonzero ($\partial h / \partial t \neq 0$) to preserve the spherical-cap geometry. By continuity, this requires a negative flow rate in the region where evaporation is arrested but the interface still recedes, i.e. an inward flow from the contact line towards the center. This produces a reverse advective protein flux that enhances rim thickening (see figure~\ref{fig:MecRing}b). However, this reverse flow has not been observed in experiments with identical droplet compositions using particle tracking velocimetry \citep{martinezpuig2025}.  

The \textit{squeezing effect} could be suppressed by incorporating a viscosity that increases in a steep way with the protein concentration. Such a dependence is supported by experimental evidence. In fact, viscosity can increase more than four orders of magnitude \cite{BansilViscosityMucin}, making the product $\mu \, Ca$ at least of order unity. Under these conditions, the droplet interface no longer needs to strictly follow a spherical-cap shape, allowing $\partial h / \partial t$ to become negligibly small. As a result, the protein ring can effectively \textit{freeze} with no internal velocities. This would suppress the fast rim broadening driven by the \textit{squeezing effect}. Unfortunately, implementing such a mechanism is not straightforward. A sharp viscosity transition generates steep gradients, which are numerically challenging and can also introduce physical instabilities. In particular, the viscosity jump could induce a local negative curvature, promoting droplet breakup—an effect not observed in our experiments. It is worth mentioning here that negative curvatures, and early touch-down of the droplet interface is indeed observed in drops of colloidal suspensions. This effect has been reproduced theoretically by \cite{coombs2024colloidal} by introducing a concentration-dependent effective viscosity.

Finally, a few words are in place about the validity of the lubrication approximation used to derive the equations in section \ref{sec:RedModel}. \cite{hu2005analysis} show a reasonably good agreement between the velocity field predicted by lubrication theory and that computed with a full two-dimensional, axisymmetric simulation. Thus, the lubrication approximation is not expected to give accurate results at short times for those drops with contact angles of up to 60$^\circ$. Also, the solute may not be well mixed along the vertical direction at this short times, especially far from the rim region. Despite these limitations, we deem the lubrication approximation as good enough to describe the rim formation for two reasons: first, the ring forms close to the contact line, where the drop height is always small; and second, an important part of the final ring mass accumulates at times when the contact angle is small compared with the initial one. Notice also that, at short times, light reflection near the contact line precludes us from determining the ring thickness (see section \ref{sec:Experiments}) so a quantitative comparison between experiments and theory is not attempted during that period.

\subsection{Comparison with real saliva material properties}\label{sec:salivaResults}

\begin{figure}[ht!]
    \centering
    \includegraphics[width=\textwidth]{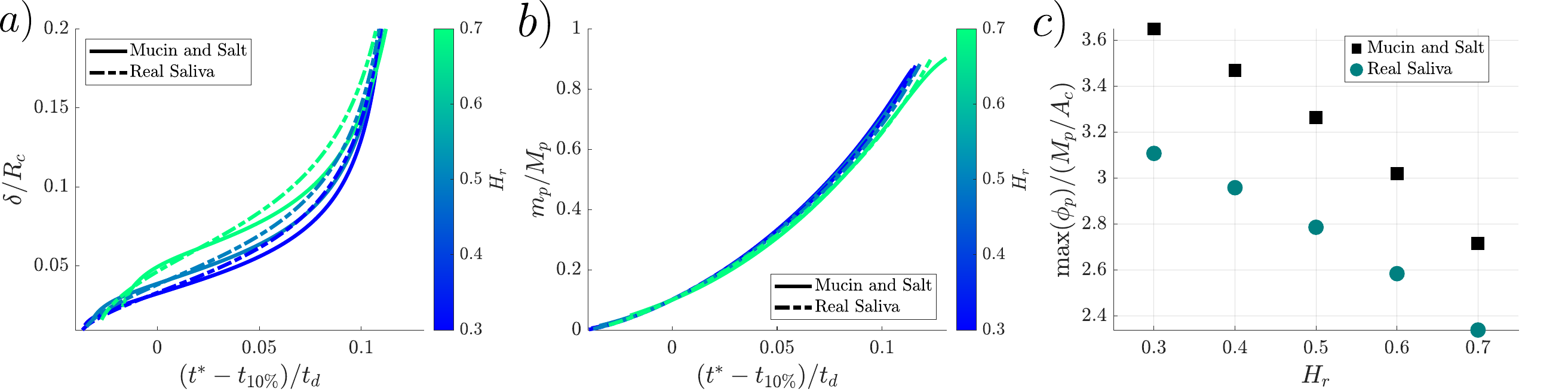}
    \caption{Comparison of simulations with real saliva material properties measured by \cite{merhi2022assessing} and the material properties used in figures~\ref{fig:Exp_vs_Sim_DeltaMass} and figures~\ref{fig:Exp_vs_Sim_PhiMaxPhi} to compare model and experiments. (a) Temporal evolution of the ring width. (b) Temporal evolution of the protein mass in the ring, normalized by the total protein mass in the droplet. (c) Maximum height-integrated protein mass fraction, normalized by the total protein mass in the droplet divided by the contact area $A_c$, as a function of relative humidity. The parameters used are $Ca = 10^{-4}$, $Pe = 43.3(1 - H_r)$, initial protein mass fraction $w_{p_0} = 3 \cdot 10^{-3}$ for mucin and salt droplets and $w_{p_0} = 12 \cdot 10^{-3}$ for real saliva droplets. Simulations are stopped at $t=t_{ev,w}$.}
    \label{fig:Num_vs_Num_Mehri}
\end{figure}

After establishing that our model agrees well with experiments when a quasi-ideal water activity is assumed, a natural question is how the predictions compare when using physicochemical properties measured for real saliva. This issue is particularly relevant for rationalizing experimental findings on sessile respiratory-like droplets used to assess viral infectivity under different environmental conditions \citep{kong2022virus,pan2025mucin}.  
A substantial body of literature reports a strong dependence of viral infectivity on relative humidity inside evaporating droplets \citep{longest2024review}. From a fluid-mechanics perspective, one of the main differences between droplets evaporating at different relative humidities is the morphology of the protein rim \citep{kong2022virus,pan2025mucin}, a trend that we also confirm experimentally (see Section~\ref{sec:Experiments}).  
To assess whether our model can be extended to real saliva, we have carried out additional simulations using the diffusivity and water-activity data reported by \citet{merhi2022assessing} (see Section~\ref{sec:WatActMatProp} for details on the implementation of these data). As shown in Fig.~\ref{fig:Num_vs_Num_Mehri}, the main experimental features are well reproduced when real saliva properties are used, namely: (a) a faster growth rate of the protein rim (in non-dimensional time) at high relative humidity (see figure~\ref{fig:Num_vs_Num_Mehri}a), and (b) a decrease in the maximum protein concentration with increasing relative humidity (see figure~\ref{fig:Num_vs_Num_Mehri}c).  
Some differences nevertheless arise when comparing simulations using real saliva properties with those assuming quasi-ideal water activity. In particular, the maximum protein concentration in the rim is smaller when real saliva data is used. This follows naturally from deviations from ideality in the water activity, which require a larger amount of protein to satisfy the condition $\chi_w \to H_r$. At the same time, the relative variation of the maximum protein concentration with relative humidity is reduced. Although the maximum protein concentration differs, Fig.~\ref{fig:Num_vs_Num_Mehri}b shows that the total mass of the protein rim grows similarly across relative humidities and remains comparable to the quasi-ideal case. As a result, since the rim mass evolves similarly while the peak concentration is lower, the protein rim is narrower when real saliva properties are used (see figure~\ref{fig:Num_vs_Num_Mehri}a).


\section{Discussion}\label{sec:Discussion}

The coffee-ring effect is one of the most widely studied phenomena in sessile evaporating droplets. The seminal works of \cite{deegan1997cofrin, deegan2000contact} brought this effect to the forefront of fluid mechanics research. Since then, more advanced theoretical models have been developed \citep{Popov, Kaplan_Mahadevan_2015, moore, Moore_ArbCon, WilsonSing,sprittles}, all focused on particle-laden droplets. These models typically assume that the evaporation rate is decoupled from solute transport. However, as we have demonstrated in this work, such decoupling is not appropriate for multicomponent droplets, where water activity depends on solute concentration, which significantly influences evaporation dynamics.

The widespread recognition of the coffee-ring effect has led to its application in interpreting experimental observations in complex fluids. For instance, \cite{kong2022virus} investigated droplets containing proteins and salts, analyzing their results within the classical particle-based framework. Yet, they reported a linear dependence of ring width on relative humidity—a finding inconsistent with theoretical predictions which suggest that the ring width is independent of humidity. Similarly, \cite{pan2025mucin} studied model respiratory droplets with compositions similar to those considered in our work. They also observed a strong relative humidity dependence in both the width and intensity of the protein ring but still interpreted their findings through the lens of particle-laden droplet theory.

Our results provide a natural explanation for these discrepancies. By incorporating a realistic coupling between hydrodynamics and solute transport through an evaporation rate that depends on the local water activity, our minimal model captures key experimental features of complex droplet evaporation. In particular, the dependence of $\chi_w$ on solute concentration alone is sufficient to qualitatively reproduce the observed humidity dependence of ring width and intensity. This underscores the importance of including water activity effects in models of complex fluids to accurately predict deposition patterns. Notably, while the classical theory of \citet{Popov} reproduces the temporal evolution of the ring mass reasonably well, it fails to capture the internal ring structure, including both its thickness and intensity. Some recent models that include particle jamming \citep{coombs2024colloidal} or particle-substrate adhesion \citep{d2025effect} are also able to describe the ring structure for drops of colloidal suspensions, although they do not predict the observed dependence of the ring structure on the relative humidity, since they still assume constant solvent activity.

Accounting for variations in the protein ring structure is also highly relevant in the context of viral infectivity. Relative humidity is known to influence virus viability in aerosol and droplet residues \citep{morris2021mechanistic, longest2024review}. In evaporating sessile droplets, this effect must be understood in terms of both the drying dynamics and the composition of the final deposit. Experiments have shown that lower relative humidity leads to thinner protein rings, while higher humidity results in lower peak protein concentrations. Given that virions have been observed to colocalize with protein-rich regions in dried droplets \citep{pan2025mucin, martinezpuig2025}, it is plausible that variations in the protein ring—driven by relative humidity—play a role in modulating viral decay. The potential impact of this variations on viral infectivity remain to be explored further by biologists. However, what can be stated a priori is that, if the relative humidity at which droplets evaporate plays a role in viral infectivity—given that $H_r$ is primarily a physical quantity that does not directly affect the infectivity of individual virions, but rather acts through differences in the evaporation dynamics within droplets—then the dependence of protein rim morphology on relative humidity, through the role of water activity, provides, from our perspective, a physically consistent framework to interpret such effects.

In conclusion, although our model is intentionally minimal, it provides a robust framework for capturing key experimental behaviors in complex evaporating droplets. By incorporating water activity–dependent evaporation, we account for the essential coupling between hydrodynamics and solute transport that is overlooked in classical models. Although we focus on model respiratory droplets the model may be extendable to other complex fluid. We hope this work stimulates further theoretical developments on the coffee-ring effect in multicomponent and biologically relevant systems. A more comprehensive two-dimensional models incorporating additional effects such as Marangoni flows and natural convection have been developed elsewhere \citep{martinezpuig2025}, but there remains significant potential to extend these approaches across a wider range of compositions and conditions.

\section*{Acknowledgements}

We are grateful to David Díaz González, Israel Pina García and Manuel Santos Rodríguez for building the humidity-controlled chamber used in the experiments. We also thank Aránzazu de la Encina, Teresa Bartolomé, Marta Sanz and Fernado Usera for their careful preparation of the protein and salt solutions used in the experiments.

\section*{Funding}
The authors acknowledge financial support from Grant No. PID2023-146809OB-I00 funded by MICIU/AEI/10.13039/501100011033 and by ERDF/UE and Grant No. PID2020-114945RB-C21 funded by MCIN/AEI/10.13039/501100011033.

\section*{Declaration of Interests}
The authors report no conflict of interest.


\section*{Final deposition state for different relative humidity}\label{app:Exp_Residue_by_Hr}

Figure~\ref{fig:ResiduePrior} shows the last frame analyzed of the evaporation process for droplets evaporated at three different ambient conditions. These images show that the rim is thicker at higher relative humidity, while the contrast between the protein rim and the center of the droplet is smaller, indicating a reduced relative protein concentration in the rim compared to the bulk. Both features agree qualitatively with the thickness and maximum protein concentration data shown in figure~\ref{fig:ResExp}.

    \begin{figure}[ht!]
    \centering
    \includegraphics[width=\textwidth]{Figures/ResiduePriorToCrys.pdf}
    \caption{Last frame analysed of the evaporation process for three droplets evaporated under different ambient conditions: a) $H_r = 30.4\%$ and $T =20.5^\circ$C; b) $H_r =44.9\%$ and $T =23.6^\circ$C; c) $H_r =67.4$ and $T =25.7^\circ$C.}
    \label{fig:ResiduePrior}
\end{figure}

\section*{Fitting of experimental material properties}
\label{app:FitExperimentalData}

For both the water activity dependence on mucin concentration, $\chi_w^p$, reported by \cite{znamenskaya}, and the diffusivity data for real saliva reported by \cite{merhi2022assessing}, we fitted analytical functional forms proposed by \cite{Salmon} to the experimental data.

\begin{figure}[ht!]
    \centering
    \includegraphics[width=\textwidth]{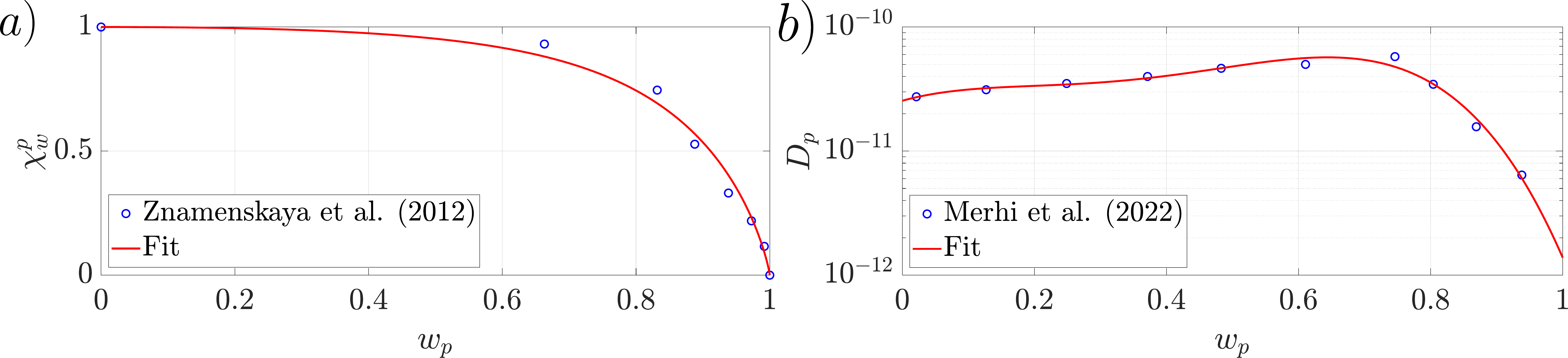}
    \caption{(a) Fitted water activity data from \cite{znamenskaya} into the functional form proposed by \cite{Salmon}. (b) Fitted diffusivity data from \cite{merhi2022assessing} into the functional form proposed by \cite{Salmon}.}
    \label{fig:Diff_WAct_Fit}
\end{figure}

To perform the fitting, we used an in-house MATLAB code to manually extract experimental data points from the original published figures. These data were then used to fit the chosen functional forms by adjusting the corresponding free parameters. As shown in Fig.~\ref{fig:Diff_WAct_Fit}, the fitted curves reproduce the experimental data well in both cases and yield reasonable extrapolations outside the range of reported measurements. This further justifies the use of the functional forms introduced in Section~\ref{sec:WatActMatProp}.

\section*{Relevance of Ideal vs. Non-Ideal Water Activity}\label{app:AwB}

\begin{figure}[ht!]
    \centering
    \includegraphics[width=\textwidth]{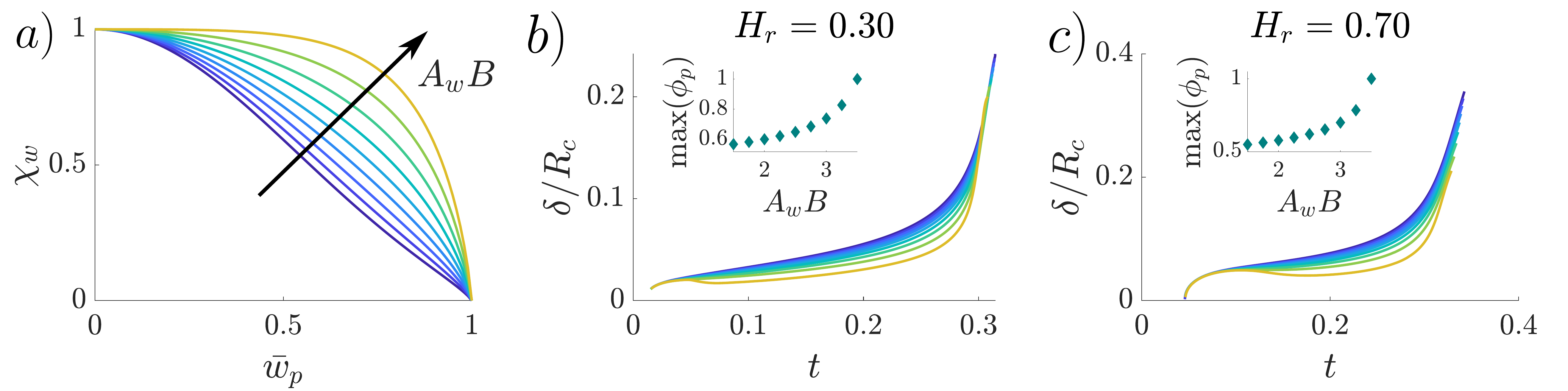}
    \caption{Effect of mixture non-ideality on protein ring morphology. (a) Water activity as a function of protein mass fraction for different values of the non-ideality parameter $A_{wB}$ ranging from 1.5 to 3.5 in steps of 0.25. (b) Temporal evolution of the protein ring width at relative humidity $H_r = 0.3$ for the same $A_{wB}$ values shown in (a). The inset displays the maximum height-integrated protein fraction as a function of $A_{wB}$. (c) Protein ring width evolution at higher humidity, $H_r = 0.7$, illustrating the impact of humidity on ring morphology under similar non-ideality conditions.}
    \label{fig:AppAwB}
\end{figure}

\cite{merhi2022assessing} demonstrated the importance of considering non-ideal water activity in analyzing the equilibrium shape of a spherical respiratory droplet. In an analogous manner, we investigate how the coffee-ring effect is influenced by the functional form of water activity as a function of protein concentration, adopting the empirical relation proposed by \cite{Salmon}:

\begin{equation*}
\chi_w = (1 - \bar{w}_p)e^{\bar{w}_p + \xi \bar{w}_p^2},
\end{equation*}

where $\xi = A_{wB} - 3(1 - \bar{w}_p)^{0.09}$. This formulation allows us to tune the non-ideality of the solution through the parameter $A_{wB}$. When $A_{wB} = 1.5$, the water activity closely approximates that of an ideal mixture, i.e., $\chi_w \approx 1 - \bar{w}_p$. For $A_{wB} = 3.5$, the system becomes strongly non-ideal, with water activity remaining nearly constant at low protein concentrations and dropping sharply only at high concentrations (see figure~\ref{fig:AppAwB}a).

In ideal mixtures, strong differences in ring formation are expected for varying relative humidity ($H_r$), since the condition $\chi_w(\bar{w}_{p,c}) \approx H_r$ is satisfied at significantly different protein concentrations. This is confirmed by our simulations: for an ideal case ($A_{wB} = 1.5$, dark blue lines in figures~\ref{fig:AppAwB}b and c), the final ring width is about 20\% of the droplet radius at $H_r = 0.3$, and nearly 40\% at $H_r = 0.7$. In contrast, in the non-ideal case ($A_{wB} = 3.5$), the final ring width is nearly independent of $H_r$.

The impact of water activity shape is particularly pronounced at high humidity. In an ideal mixture, $\chi_w(\bar{w}_{p,c}) \approx 0.7$ occurs at $\bar{w}_{p,c} \approx 0.3$, whereas for a strongly non-ideal mixture, this condition is only satisfied when $\bar{w}{p,c}$ is close to 1. At low humidity, this difference is less significant, since even ideal mixtures reach $\bar{w}_{p,c} \approx 1$.

\section*{Evaporation rate}\label{app:EvRate}

To compute the evaporation rate, we first represent the water activity $\chi_w(r)$ using a cubic spline interpolation. This representation introduces two degrees of freedom at the boundaries of the domain, which must be constrained by appropriate boundary conditions. To ensure physical consistency, we impose symmetry conditions at the droplet center

\begin{equation*}
    \frac{\partial \chi_w}{\partial r} = 0 , \  \mathrm{at} \ r=0.
\end{equation*}
At the contact line, the derivative is determined by applying the chain rule

\begin{equation*}
    \frac{\partial \chi_w}{\partial r} = \frac{\partial \chi_w}{\partial w_p} \frac{\partial w_p}{\partial r}, \  \mathrm{at} \ r=1.
\end{equation*}

\begin{figure}[ht!]
    \centering
    \includegraphics[width=\textwidth]{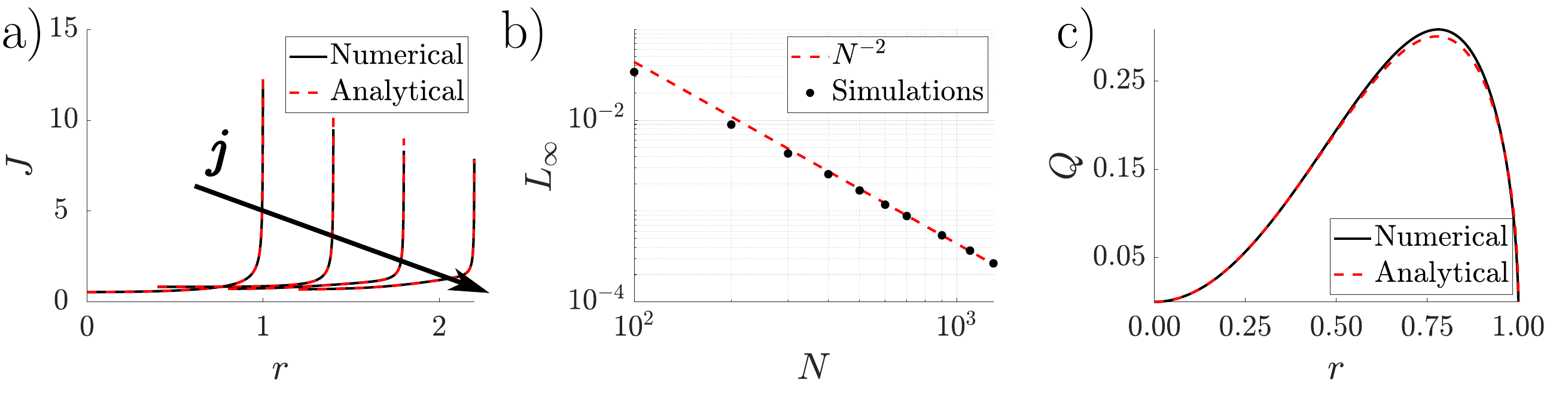}
    \caption{Validation of the numerical framework. (a) Comparison of the semi-analytical evaporation rate with the analytical solution from \cite{FUDopantConcentration} for different polynomial orders $j = 0, 2, 4, 6$. Each curve is shifted by 0.5 along the horizontal axis for clarity. (b) Evolution of the error $L_{\infty}$, defined as the maximum deviation from mass conservation during the simulation, plotted against the number of grid points. (c) Comparison of the flow rate predicted by our model with the analytical solution from \cite{gelderblom2022evaporation}, using parameters $Ca = 10^{-4}$, $Pe = 10$, $H_r = 0.5$, and at time $t = 0.1$, early enough to assume $\chi_w \approx 1$.}
    \label{fig:AppNum}
\end{figure}

Given that the cubic splines are formed by cubic polynomials defined between grid points where the value of $\chi_w$ is known, the spline $S\chi_w$ can be written as

\begin{equation*}
    S\chi_w(r) = \sum_{j=0}^3 a_j^i r^j, \ \mathrm{for} \ r\in(r_i, r_{i+1}),
\end{equation*}
where $a^i_j$ is the spline coefficient of the polynomial of order $j$ between grid points $i$ and $i+1$. We can then express $g(r')$ as

\begin{equation*}
    g(r') = \frac{2}{\pi}\frac{d}{dr'}\left(\sum_{i=0}^{L}\sum_{j=0}^3\int_{r_i}^{r_{i+1}}\frac{a_j^ir^{j+1}}{\sqrt{r'^2-r^2}}dr + \sum_{j=0}^3\int_{r_{L+1}}^{r'}\frac{a_{L+1}^ir^{j+1}}{\sqrt{r'^2-r^2}}dr\right),
\end{equation*}
where $L$ is the maximum grid-point such that $r_{L+1}<r'$. This formulation is particularly convenient, as the integrals involved in computing g can be evaluated analytically. Once $g$ is obtained , the evaporation rate $J$ can be computed through numerical integration. For this purpose, we employ Gauss-Kronrod quadrature, as implemented in the \texttt{quadgk} function in MATLAB\textsuperscript{\textregistered}. We verify our implementation by comparing the semi-analytical results with the analytical solution provided by~\cite{FUDopantConcentration} for the case in which the water activity $\chi_w$ is given by a polynomial function. In figure \ref{fig:AppNum}a, we compare the analytical and numerical solutions for $\chi_w = 1 - 0.15\cdot r^j$, with $j = \{0, 2, 4, 6\}$. It is important to note that, to ensure axisymmetry of the solution, the polynomial order must be even.

\section*{Numerical scheme}\label{app:NumSch}

Due to the high non-linearity of the thin film equation \eqref{eq4:NonDimensionalModel} the spatial discretization must be done carefully. Equations \eqref{eq4:NonDimensionalModel} can be written as

$$
\frac{\partial(r h)}{\partial t}+\bar{u}_r \frac{\partial(r h)}{\partial r}+r h \frac{\partial \bar{u}_r}{\partial r}=J,
$$

We employ a numerical scheme similar to that used by \citet{StoneCapHealing}, based on two uniform staggered grids: one defined at the primary nodes $r_i$, and the other at the midpoints $r_{i+1/2}$. A second-order finite difference scheme is applied to approximate spatial derivatives,

$$
\frac{\partial h_i}{\partial t}=-\frac{\bar{u}_i }{r_i} \frac{r_{i+1} h_{i+1}-r_i h_i}{r_{i+1}-r_i}-h_i \frac{\bar{u}_{i+1 / 2}-\bar{u}_{i-1 / 2}}{\left(r_{i+1}-r_{i-1}\right) / 2},
$$
where
$$
\begin{aligned}
\bar{u}_i  & =\frac{\left(h_i \right)^2}{3\mu_i Ca} \frac{p_{i+1} -p_{i-1} }{r_{i+1}-r_{i-1}} \\
\bar{u}_{i+\frac{1}{2}}  & =\frac{1}{3 Ca \left(\mu_{i+1}+\mu_i\right)/2} \left(\frac{h_{i+1} +h_i }{2}\right)^2 \frac{p_{i+1} -p_i }{r_{i+1}-r_i} \\
p_i  & =\frac{2}{r_{i+1}-r_{i-1}}\left(\frac{h_{i+1} -h_i }{r_{i+1}-r_i}-\frac{h_i -h_{i-1} }{r_i-r_{i-1}}\right)+\frac{1}{r_i} \frac{h_{i+1} -h_{i-1} }{r_{i+1}-r_{i-1}}.
\end{aligned}
$$

To discretize spatially the advection-diffusion equation we first defined, following \cite{moore}, the integrated mass variable 

\begin{equation*}
    \Phi_p (r,t) = \int_0^r r'\phi_p(r',t)dr'.
\end{equation*}
In terms of this integrated variable the advection-diffusion equation \eqref{eq4:SoluteDimensionaless} becomes

\begin{equation*}
\frac{\partial \Phi_p }{\partial t}+\left(\bar{u}_r + \frac{1}{Pe}\left(\frac{D_p}{r}+\frac{D_p}{h}\frac{\partial h}{\partial r}\right)\right)\frac{\partial \Phi_p }{\partial r}=\frac{1}{Pe}D_p\frac{\partial^2 \Phi_p }{\partial r^2}.
\end{equation*}
with boundary conditions

\begin{equation*}
    \Phi_p(0,t)=0, \ \Phi_p(1,t)=\frac{1}{4},
\end{equation*}
and initial condition
\begin{equation*}
    \Phi_p(r,0)=\frac{r^2}{2}-\frac{r^4}{4}.
\end{equation*}
This equation is discretized using a second-order finite difference scheme at the primary grid nodes $r_i$. The resulting spatially discretized system of coupled hydrodynamic and transport equations is integrated in time using \texttt{ode15s} in \textsc{MATLAB}\textsuperscript{\textregistered}, with a relative tolerance of $10^{-4}$ and an absolute tolerance of $10^{-7}$.We verify the expected second-order spatial convergence of the method through a representative example shown in figure \ref{fig:AppNum}b, for a Péclet number $Pe = 30$ and a capillary number $Ca = 10^{-4}$. In addition, we validate the hydrodynamic implementation by comparing the computed flow rate with the analytical solution reported by \cite{gelderblom2022evaporation}. This comparison is performed at early times on the evaporation, when the water activity satisfies $\chi_w \approx 1$ (see figure \ref{fig:AppNum}c).

\section*{Estimation of gravity effects on the mean radial velocity}\label{app:BuoyancyDrivenFlow}

Using the Boussinesq approximation, the radial and axial components of the momentum conservation equation can be written \citep{Diddens2021JFM}:
\begin{equation}
0 = -\partial_r P + \mu \frac{\partial}{\partial r}\left(\frac{1}{r}\frac{\partial (rv_r)}{\partial r}\right) + \mu \frac{\partial^2v_r}{\partial z^2}.
\end{equation}
\begin{equation}
0 = -\partial_z P + \mu \frac{1}{r}\frac{\partial}{\partial r}\left(r\frac{\partial v_z}{\partial r}\right) + \mu \frac{\partial^2v_z}{\partial z^2} - \partial_w\rho w g.
\end{equation}
Here, $P$ is the reduced pressure and $\partial_w\rho$ is the variation of the mixture density with the solute mass fraction $w$. For the sake of the argument, we restrict our analysis here to Stokes flow with constant viscosity $\mu$ and diffusivity $D$. From the $z$-component of the momentum conservation equation, we infer that the order of magnitude of the pressure variations along the $z$ direction are:
\begin{equation}
\Delta_z P \sim \partial_w\rho \Delta w g h_c,
\end{equation}
where $\partial_w\rho$ is the derivative of the fluid density with respect to the solute mass fraction $w$, and $\Delta w$ is the characteristic variation of the solute mass fraction along the vertical extension of the drop, $h_c$. Introducing these pressure variation into the radial component of the momentum conservation equation:
\begin{equation}
v_{r, g} \sim \frac{\partial_w\rho \Delta w g h_c^3}{\mu R_c},
\end{equation}
where we have assumed that the radial pressure variations are also of order $\Delta_zP$. This makes sense, as both the mass fraction and the drop height change of the order of themselves along the drop radius. To determine the typical variation of the mass fraction along the drop's height we look at the boundary condition that imposes no flux of solute across the free surface,
\begin{equation}
w \left(\vec{v}-\vec{v}_I\right)\cdot\vec{n} - D \nabla w \cdot \vec{n} = 0.
\end{equation}
Here, $\vec{v}_I$ is the velocity of the interface. To simplify the argument we further assume $\epsilon = h_c/R_c \ll 1$. In these conditions, this boundary condition projects on the vertical axis, thus
\begin{equation}
w (v_z-v_{I, z}) - D_p\frac{\partial w}{\partial z} \approx 0.
\end{equation}
The difference between the liquid $\vec{v}$ and the interfacial $\vec{v}_I$ velocity at the interface is of the order of $J/\rho$, where $J$ is the local evaporative mass flux per unit area of interface. Thus,
\begin{equation}
\Delta w \sim \frac{J_c}{\rho} \frac{h_c w_c}{D_p},
\end{equation}
where $J_c$ and $w_c$ are the characteristic evaporative flux and solute mass fraction, respectively. Introducing this into the estimation of the radial velocity induced by buoyancy,
\begin{equation}
v_{r, g} \sim \frac{J_c}{\rho} \left(\frac{h_c}{R_c}\right)^4 \frac{\partial_w\rho g w_c R_c^3}{\mu D_p} = \frac{J_c}{\rho} \epsilon^4 \, \mathrm{Ra} \, w_c,
\end{equation}
where we define a Rayleigh number as $\mathrm{Ra} = \partial_w\rho\,g\,R_c^3/\mu D_p$. The turnover time of the gravity-driven recirculation can then be expressed as
\begin{equation}
    t_\mathrm{Ra} \sim R_c / v_{r, g} \sim \frac{\rho R_c}{J_c\,\epsilon^4\,\mathrm{Ra}\,w_c}.
\end{equation}
This time scale must be compared to the typical evaporation time,
\begin{equation}
    t_\mathrm{ev} \sim \frac{R_c^2 h_c}{(J_c/\rho) R_c^2} \sim \frac{\epsilon\,R_c}{J_c/\rho}.
\end{equation}
Finally, the ratio between the buoyancy-driven turnover time and the evaporation time is
\begin{equation}
\frac{t_\mathrm{Ra}}{t_\mathrm{ev}} \sim (\mathrm{Ra}\,w_c)^{-1}\,\epsilon^{-5}.
\end{equation}
To evaluate this ratio we take, as a typical situation, the model respiratory droplets studied in \cite{martinezpuig2025}, where $\mathrm{Ra} \sim 10^6$ and $w_c \approx 3\times10^{-3}$. On the other hand, for a typical contact angle $\theta = 45^\circ$, we have $\epsilon = (1-\cos\theta)/\sin\theta \approx 0.414$. With these data
\begin{equation}
    \frac{t_\mathrm{Ra}}{t_\mathrm{ev}} \approx 1/37.
\end{equation}
As pointed out by Diddens and co-workers, this low value of the time ratio means that the drop volume stays nearly constant during the time a particle takes to complete a turn. Thus, the buoyancy-driven recirculatory motion can be modeled as quasi-steady. A consequence of this is that the normal velocity to the interface induced by the buoyancy-driven flow is very small which, in turn, imposes that the recirculatory flow cannot transport a net mass flux. Therefore, the rim is formed by the transport done by the evaporation-driven capillary flow, as we model in our manuscript.

We must highlight that this estimation is somewhat conservative, as in fact the velocities near the rim are going to be larger than the typical value estimated assuming a uniform evaporative flux $J_c$. Moreover, the velocity induced by the recirculation is going to be smaller than the value estimated above close to the rim region, where the drop is the shallowest. All in all, the buoyancy-driven flow does not induce a significant net transport towards the rim and can be ignored to study the rim dynamics.

\clearpage

\bibliographystyle{jfm}
\bibliography{main}



\end{document}